\documentclass[a4paper]{article}
\input{topmatter}

\newcommand{\Rho}{\mathrm{P}}

\def\BB{{\cal B}}
\def\CC{{\cal C}}
\def\DD{{\cal D}}

\def\HH{{\cal H}}
\def\II{{\cal I}}
\def\JJ{{\cal J}}
\def\KK{{\cal K}}
\def\LL{{\cal L}}
\def\MM{{\cal M}}
\def\NN{{\cal N}}
\def\OO{{\cal O}}
\def\PP{{\cal P}}
\def\QQ{{\cal Q}}
\def\RR{{\cal R}}

\def\det{{\rm det}}

\def\dim{{\rm dim}}

\def\ch{{\rm ch}}

\def \cf{{\it cf.}}

\newcommand{\be}{\begin{equation}}
\newcommand{\ee}{\end{equation}}
\newcommand{\bea}{\begin{eqnarray}}
\newcommand{\eea}{\end{eqnarray}}
\newcommand\zb{\bar{z}}

\def\aa{\alpha}
\def\aad{{\dot{\alpha}}}
\def\bb{\beta}
\def\bbd{{\dot{\beta}}}
\def\ggd{{\dot{\gamma}}}

\def\ddd{\dot{\delta}}
\def\ii{i}
\def\Lb{\bar{L}}
\def\pd{\dot{+}}
\def\md{\dot{-}}
\def\Lh{\hat{L}}

\def\reg{\mathrm{regular}}
\def\dagplane{*}
\def\dagrad{\dagger}

\def\zb{\bar{z}}

\newcommand\qq{\mathbbmtt{Q}}

\newcommand{\upsideDown}[1]{%
  \reflectbox{\rotatebox[origin=c]{180}{#1}}%
}

\newcommand{\Rdec}{\upsideDown{$\mathfrak{R}$}}

\newtheorem{remark}{Remark}

\title{Graded Unitarity in the SCFT/VOA Correspondence}

\author[a]{Arash Arabi Ardehali,}
\author[b]{Christopher Beem,}
\author[c]{Madalena Lemos,}
\author[a]{Leonardo Rastelli}

\affiliation[a]{C.N. Yang Institute for Theoretical Physics, Stony Brook University, Stony Brook, NY 11794, USA}
\affiliation[b]{Mathematical Institute, University of Oxford, Woodstock Road, Oxford, OX2 6GG, UK}
\affiliation[c]{Department of Mathematical Sciences, Durham University, Upper Mountjoy, DH1 3LE Durham, UK}
\abstract{Vertex algebras that arise from four-dimensional, $\mathcal{N}=2$ superconformal field theories inherit a collection of novel structural properties from their four-dimensional ancestors. Crucially, when the parent SCFT is unitary, the corresponding vertex algebra is not unitary in the conventional sense. In this paper, we motivate and define a generalized notion of unitarity for vertex algebras that we call \emph{graded unitarity}, and which captures the consequences of four-dimensional unitarity under this correspondence. We also take the first steps towards a classification program for graded-unitary vertex algebras whose underlying vertex algebras are Virasoro or affine Kac--Moody vertex algebras. Remarkably, under certain natural assumptions about the $\mathfrak{R}$-filtration for these vertex algebras, we show that only the $(2,p)$ central charges for Virasoro VOAs and boundary admissible levels for $\mathfrak{sl}_2$ and $\mathfrak{sl}_3$ Kac--Moody vertex algebras can possibly be compatible with graded unitarity. These are precisely the cases of these vertex algebras that are known to arise from four dimensions.}

\begin{document}
\maketitle
\flushbottom


\section{\label{sec:intro}Introduction}

To any four-dimensional ${\cal N}=2$ superconformal field theory (SCFT) one canonically associates~\cite{Beem:2013sza} a vertex operator algebra (VOA):
\begin{equation}\label{VV}
    \mathbb{V}: \quad {\rm 4d\; SCFT\; \rightarrow VOA}~.
\end{equation}
The VOA arises as the cohomological reduction of the full local operator product expansion (OPE) algebra of the four-dimensional theory ${\cal T}$ with respect to a certain nilpotent supercharge. As a vector space, the VOA $\mathbb{V}[{\cal T}]$ comprises the \emph{Schur operators} of ${\cal T}$, a class of local operators belonging to specific shortened representations of the four-dimensional ${\cal N}=2$ superconformal algebra. 

The map $\mathbb{V}$ appears to be essentially injective, as no examples are known of two genuinely distinct four-dimensional theories that yield the same VOA.\footnote{Two caveats are in order for this statement to be true. First, we work modulo exactly marginal deformations, as theories on the same conformal manifold map to the same VOA. Second, we only keep track of the \emph{local} operator algebra of the theory in $\mathbb{R}^4$. There are well-known examples of theories which have the same local OPE algebra but differ in their extended operator data or when placed on spaces with nontrivial topology (\emph{i.e.},~Lagrangian theories with different choices of compact gauge group for given gauge algebra). It is generally believed that the finer classification that includes such ``global'' data amounts to at most a finite number of additional choices for each local OPE algebra.} However, it is not surjective. According to a basic entry of the 4d/2d dictionary, the VOA central charge $c$ is related to the four-dimensional Weyl anomaly coefficient $c_{4d}$ by the universal relation $c= -12 c_{4d}$. It follows that any unitary four-dimensional SCFT (for which $c_{4d} >0$) maps to a nonunitary VOA. More dramatically, a central conjecture~\cite{Beem:2017ooy}, for which there is now overwhelming evidence, asserts that the Higgs branch ${\cal M}_H$ of the four-dimensional theory (viewed as a holomorphic symplectic variety) is isomorphic with the associated variety \cite{Arakawa:2010ni} of the corresponding vertex algebra,
\begin{equation}
    {\cal M}_H[{\cal T}]= X_{{\mathbb V}[{\cal T}]}~.
\end{equation}
Granting this conjecture, VOAs that arise from this correspondence must be \emph{quasi-lisse}~\cite{Arakawa2018}, which by definition means that their associated varieties have a finite number of symplectic leaves. Quasi-lisse VOAs enjoy interesting modular properties, partly generalizing those of rational VOAs. To wit, their vacuum character~\footnote{In fact, the character of any simple ordinary module. A quasi-lisse VOA has finitely many simple ordinary modules~\cite{Arakawa2018}.} (which in the 4d/2d dictionary corresponds to the Schur index of the parent theory) obeys a finite-order, weight-zero, modular linear differential equation (MLDE). The connection with the physics of the Higgs branch goes further. At least in a large class of examples, one can construct remarkably well-behaved free field realizations for $\mathbb{V}[{\cal T}]$, whose ingredients can be read off from physical data on the Higgs branch of ${\cal T}$ \cite{Beem:2019tfp,Beem:2019snk,Beem:2021jnm,Beem:2024fom}.

It is clear that VOAs in the image of the map $\mathbb{V}$ enjoy structural features that make them better behaved than generic VOAs in many ways. It would be desirable to distill the properties of VOAs that descend from four dimensions into a set of additional axioms (on top of the standard axioms of a VOA) to allow for their intrinsic study without reference to four-dimensional physics. Their classification, for example, could then be made into a well-posed mathematical problem. The principal aim of this work is to take steps towards such an axiomatization.

We start with the vector space underlying a VOA in the image of $\mathbb{V}$. This is identified with the vector space ${\cal V}$ of Schur operators in the pre-image theory, and therefore comes equipped with a triple grading. One grading is by the $U(1)_r$ charge, which plays the role of cohomological degree in the reduction that yields the VOA from the parent SCFT; we will denote it by a superscript, ${\cal V}^\bullet$. The other two gradings are by the holomorphic conformal weight $h$ and by the (half-integral) weight $R$ of the Cartan subalgebra of the $\mathfrak{su}(2)_R$ symmetry, so we have
\begin{equation}
    \mathcal{V} = \bigoplus_{h,R\in\frac12\mathbb{N}}\mathcal{V}^{\,\bullet}_{h,R}~.
\end{equation}
While conformal weight and cohomological degree fit naturally into the standard axiomatic package of a VOA, the additional $R$ quantum number is not and plays a central role in structuring the VOAs in the image of $\mathbb{V}$.

For example, for a given conformal weight $h$, four-dimensional considerations dictate that all states must obey $R\leqslant h$. The value of $R$ keeps track of the precise four-dimensional superconformal multiplet to which a given Schur operator belongs, which is precious physical information. Operators with $R=h$ (the maximum possible value at a fixed conformal weight) are the \emph{Higgs branch chiral ring operators}, which have nonsingular operator products and define a commutative $\mathbb{C}$-algebra that is identified with the coordinate ring $\mathbb{C}[{\cal M}_H]$ of the Higgs branch. 

What makes the $R$-grading subtle relative to the associated VOA is that the vertex algebra OPE is not $R$-equivariant. However, defining instead an ascending \emph{filtration} by $R$ (which we denote by $\mathfrak{R}_\bullet\mathbb{V}$), one sees that the filtration is necessarily \emph{good} (in the technical sense). By a standard construction~\cite{li2004vertex} one can take a commutative limit of the VOA by passing to the associated graded of any good filtration; the resulting algebraic structure is that of a vertex Poisson algebra (VPA). The VPA obtained by taking the commutative limit with respect to the $\mathfrak{R}$-filtration coincides with the VPA defined by the holomorphic-topological (HT) twist~\cite{Kapustin:2006hi} of the four-dimensional parent theory \cite{Beem:2017ooy}. Conversely, one can view the VOA $\mathbb{V}[{\cal T}]$ as a deformation quantization of the VPA that arises from the holomorphic-topological twist of ${\cal T}$; this can be made transparent by realizing the VOA in terms of a certain Omega deformation~\cite{Oh:2019bgz,Jeong:2019pzg} of the holomorphically twisted theory. 

In this paper, we focus on a further fundamental, structural feature of VOAs in the image of $\mathbb{V}$: their compatibility with four-dimensional unitarity at the level of two-point functions. From a purely VOA perspective, this unitarity is a hidden feature, as the $R$-grading of states is required in order to formulate it. We provide a careful axiomatization of the requirements of four-dimensional unitarity on $\mathfrak{R}$-filtered VOAs, culminating in the definition of a \emph{graded unitary} vertex algebra. The essential point is that four-dimensional unitarity implies the existence of an order-four anti-linear VOA automorphism (which we call the \emph{conjugation} $\rho$); in terms of this conjugation and the $\mathfrak{R}$-filtration one may define an Hermitian form that differs from more standard VOA pairings and is required by unitarity/reflection positivity to be positive definite. These definitions are closely related to similar notions that hold in the context of hyperk\"ahler geometry.

With a formulation of graded unitarity in hand, we return to the question of surjectivity of the map $\mathbb{V}$. How restrictive are the constraints of four-dimensional unitarity? When asking this question in complete generality, one faces a basic obstacle. Given the presentation of VOA in terms of a set of strong generators and their singular OPEs, it is \emph{a priori} unclear how to determine the $\mathfrak{R}$-filtration. Even when one has a precise opinion about the $R$-charge assignments of the strong generators (as one often does: notably for the stress tensor and for generators of the Higgs chiral ring), to our knowledge there is no general principle that fixes the $R$-grading for composite operators, and indeed there are known examples where the naive charge assignment requires modification. For composites of low conformal weight, one can often resolve any ambiguity by appealing to the principle that an interacting SCFT should contain no conserved higher-spin currents, and already from that information one can derive some powerful consequences of four-dimensional unitarity~\cite{Beem:2013sza,Liendo:2015ofa, Lemos:2015orc,Beem:2018duj}, including several profound inequalities involving the central charge and the levels of affine Kac Moody (AKM) subalgebras. Nevertheless, it is clear that this only scratches the surface of a much richer set of constraints. 

To make progress, in this work we focus our attention on several simple classes of VOAs (assumed to arise from four dimensions) for which there are compelling general conjectures for the $\mathfrak{R}$-filtration. We first consider the case where the VOA is just the simple quotient of the Virasoro vertex algebra at some (negative) central charge $c$. The VOA (presumed to be in the image of $\mathbb V$) is then strongly generated by a single operator, the stress tensor, with standard self-OPE. We know that the set of such VOAs arising from four dimensions is nonempty: there exists an infinite sequence of SCFTs (the $(A_1, A_{2k})$ Argyres-Douglas theories) that maps to the Virasoro VOAs with $c = c_{2, 2k+3}$, where $c_{p, q}$ is the central charge of the $(p, q)$ Virasoro minimal model. One desideratum of our program would be to prove that these values of the central charge are the only ones compatible with four-dimensional unitarity. In the present work, we make the additional assumption that the $\mathfrak{R}$-filtration is such that the $p^{\rm th}$ filtered subspace is the one spanned by normal ordered products with at most $p$ stress tensors (allowing derivatives). (This is expected to be the correct filtration for the $(A_1, A_{2k})$ Argyres-Douglas theories \cite{Song:2016yfd,Foda:2019guo} on the basis of calculations of the Macdonald index~\cite{Gadde:2011uv}, which counts Schur operators with an extra fugacity that encodes $R$ charges.) Strikingly, with this assumption in place, we are indeed able to prove by a straightforward combinatorial argument involving the sign of the Kac determinant that \emph{any other value of the central charge would violate the sign-requirements of graded unitarity}.

The next natural class of examples are affine current algebras, starting with the easiest case of $\mathfrak{sl}_2$. We again posit a four-dimensional SCFT whose associated VOA is precisely the simple quotient of the $\mathfrak{sl}_2$ current algebra at some level $k$. There is again a compelling guess for the $\mathfrak{R}$-filtration, which now takes into account a well-known correction due to the Sugawara construction of the stress tensor. The normal ordered product of two currents in the singlet representation is proportional to the VOA stress tensor, which must be assigned $R=1$ (rather than the naive $R=2$ from adding the charges of the constituents). Under the assumption that this is the \emph{only} correction to the naive filtration based on counting charges of strong generators, we are able to show that the only values of the level that are not incompatible with graded unitarity are the Kac--Wakimoto boundary admissible levels $k = -2+\frac{2}{2n+1}$. Remarkably, those are the only levels that have been observed in the wild: they arise in the $(A_1, D_{2n+1})$ sequence of Argyres-Douglas SCFTs. 

One can begin to pursue an analogous analysis of current algebras of higher rank. A plausible proposal for the $\mathfrak{R}$-filtration now involves making corrections for all the higher ``Casimir operators'' (note that beyond the $\mathfrak{sl}_3$ case, establishing that such a filtration is \emph{good} appears nontrivial). We analyze in some detail the cases of $\mathfrak{sl}_3$ and $\mathfrak{sl}_4$. For $\mathfrak{sl}_3$ we again find that the only levels not disallowed are the boundary admissible ones $k_{3,q}=-3+\frac{3}{q}$ with $q$ coprime to $e$. For $\mathfrak{sl}_4$ we find that at our level of analysis, in addition to the boundary admissible levels, some inadmissible levels (namely $k_{2,q} = -4+\frac{2}{q}$ with $q$ odd) are also not ruled out. We expect that a more refined state-level analysis may be able to rule out these values; in particular for $k=k_{2,1}=-2$ the vertex algebra is known not to be quasi-lisse \cite{Arakawa_Moreau:2017157,arakawa2024generalized}, while the boundary admissible levels are known to arise in generalized Argyres--Douglas theories. It would be desirable to develop more streamlined combinatorial methods to impose graded unitarity constraints on general current algebras.

In summary, the constraints of four-dimensional unitarity appear to be extremely restrictive. Making plausible assumptions about the $\mathfrak{R}$-filtration of some natural infinite classes of VOAs that might arise from four-dimensional SCFTs, we have found that central charges and levels can only take very special discrete values. We should emphasize that we have only imposed an extremely limited subset of the graded unitarity constraints, namely that the {\it determinant} of the relevant Hermitian form should have positive sign on a given weight space. In principle, the whole form must be positive, a much stronger constraint. For the infinite sequences of VOAs that are known to arise from four-dimensional SCFTs (such as ${\rm Vir}_{2, 2k+3}$ and boundary admissible current algebras) we have compelling four-dimensional reasons to believe that the stronger positivity statements must be true. This leads to a set of intricate mathematical conjectures (positivity of a certain well-defined hermitian form on each of these filtered vertex algebras).

It would be desirable to show that the presumed $\mathfrak{R}$-filtrations are, in fact, fixed by the requirement of graded unitary itself. One may attempt to systematically ``bootstrap'' the $\mathfrak{R}$-filtration on subspaces of increasing conformal weight, imposing goodness of the filtration and positivity of the hermitian form. Preliminary explorations in this direction have been inconclusive, partly because a brute force analysis quickly becomes computationally too expensive, but the general idea may still be viable. Alternatively, there could be multiple $\mathfrak{R}$-filtrations compatible with the axioms, with the physically correct one selected by some additional principle. 
 
There is an important clue that the $\mathfrak{R}$-filtration may be intimately related to some generalized notion of geometry. As mentioned above, some large classes of VOAs arising from four dimensions admit free field realizations motivated by the effective field theory on the Higgs branch~\cite{Beem:2019tfp,Beem:2019snk,Beem:2021jnm,Beem:2024fom}. In these examples, $\mathbb{V} [{\cal T}]$ is realized as a subalgebra of a vertex algebra $\mathbb{V}_{\rm EFT}[\cal T]$, which comprises a set of chiral bosons encoding the Higgs branch geometry, as well as an ``irreducible'' building block  $\mathbb{V}[ {\cal T}_{\rm IR}]$ (whose associated variety is a point) corresponding to the residual IR SCFT ${\cal T}_{\rm IR}$ at a generic point of the Higgs branch. The chiral bosons carry a natural filtration inherited from the scaling symmetry on the Higgs branch, while $\mathbb{V}[ {\cal T}_{\rm IR}]$
carries its own $\mathfrak{R}$-filtration, often known from independent considerations (\emph{e.g.}, knowledge of the Macdonald index of ${\cal T}_{\rm IR}$). The resulting total filtration is conjectured to coincide with the $\mathfrak{R}$-filtration on $\mathbb{V} [{\cal T}] \subset {\cal V}_{\rm EFT}  [{\cal T}]$, and many checks have been carried out in various examples. The upshot is that free field realizations provide a partial geometric picture for the $\mathfrak{R}$-filtration. This picture is only partial because the $\mathfrak{R}$-filtration of the irreducible vertex algebra $\mathbb{V}[ {\cal T}_{\rm IR}]$  must be supplied by hand. 

\medskip

The organization of the paper is best apprehended from the table of contents. In Section~\ref{sec:axiomatics} we review some basic facts about the SCFT/VOA correspondence, with an emphasis on the role of the $\mathfrak{R}$-filtration. We then carefully axiomatize the implications four-dimensional unitarity at the level of two-point functions, culminating in the definition of a graded unitary vertex algebra. In Section~\ref{sec:virasoro} we enforce the constraints of graded unitarity on the simple quotient of the Virasoro algebra at central charge $c$, under the assumption that the $\mathfrak{R}$-filtration is the naive one, based on counting the number of stress tensors. We prove that graded unitarity can at most hold for the ``boundary'' minimal model levels $c=c_{2,2k+1}$---precisely the values realized in the $(A_1, A_{2k})$ sequence of Argyres-Douglas theories. In Section~\ref{sec:sl2_akm} we perform a similar analysis for the simple quotient of the $\mathfrak{sl}_2$ current algebra, assuming that the $\mathfrak{R}$-filtration is the natural one that corrects that naive filtration (based on counting currents) to account for the reduced $R$-charge of the Sugawara stress tensor. We prove that the level can at most take the boundary admissible values $k = -\frac{4n}{2n+1}$, which are realized in the $(A_1, D_{2n+1})$ sequence of Argyres-Douglas theories. In Section~\ref{sec:general_akm} we begin to generalize the analysis to current algebras of higher rank, studying in some detail the cases of $\mathfrak{sl}_3$ and $\mathfrak{sl}_4$, under a plausible assumption about the $\mathfrak{R}$-filtration. Three appendices complement the text with background material and some technical details.

\section{\label{sec:axiomatics}VOAs from four dimensions: review and structure}

Vertex operator (super)algebras\footnote{Henceforth we omit the modifier \emph{super} and consider the super and bosonic cases uniformly.} that arise from four-dimensional SCFTs via the twisted construction of \cite{Beem:2013sza} come equipped with a package of additional structures that go beyond those of a bare vertex operator algebra; these additional structures have been exploited to varying degrees in previous work on the subject (see, \emph{e.g.}, \cite{Lemos:2015orc,Beem:2018duj}). In this section, we formalize this package in the abstract. Many aspects of the story here run parallel to the related case of deformation quantization of Higgs branches arising in three-dimensional $\mathcal{N}=4$ SCFTs \cite{Beem:2016cbd}, which has been rigorously formalized in terms of \emph{short, positive star products} in \cite{Etingof:2019guc,Etingof:2020fls}. It would be very interesting to further investigate the specific connection between these two structures (along the lines of \cite{Dedushenko:2019mzv,Pan:2020cgc}) in greater detail.

\subsection{\label{subsec:4d_voa_review}Review of VOA constructions from 4d}

We briefly recall the construction of \cite{Beem:2013sza}. The starting point is a general four-dimensional $\mathcal{N}=2$ SCFT, and in particular its (pre-Hilbert) space $\mathcal{H}^{(loc)}$ of local operators inserted at a specified origin. This space admits a \emph{quintuple grading} induced by the action of the Cartan subalgebra of the (complexified) $\mathfrak{sl}(4|2)$ superconformal algebra. In a standard basis for this Cartan subalgebra,\footnote{See Appendix~\ref{app:conventions} for our conventions for the $\mathcal{N}=2$ superconformal algebra and quantum numbers.} we have the decomposition,
\begin{equation}
    \mathcal{H}^{(loc)} = \bigoplus_{E,j_1,j_2,R,r}\mathcal{H}^{(loc)}_{E,j_1,j_2,R,r}~.
\end{equation}
There is also a decomposition into irreducible representations of $\mathfrak{sl}(4|2)$, and the associated vertex operator algebra can be realized cohomologically with respect to either of two distinguished (families of) supercharges,
\begin{equation}
    \qq_{\,1}^{(\zeta)} = \mathcal{Q}^1_{-}+\zeta \widetilde{\mathcal{S}}^{2\dot{-}}~,\qquad \qq_{\,2}^{(\zeta)} = \widetilde{\mathcal{Q}}_{2\dot{-}}-\zeta \mathcal{S}_1^{-}~.
\end{equation}
The parameter $\zeta\in\mathbb{C}^\times$ can be rescaled by conjugating with a unitary action of $U(1)_r$ and spacetime dilatations, and can be taken without loss of generality to have any nonzero value (as was done in \cite{Beem:2013sza}) at the expense of fixing certain noncanonical choices of conventions. In what follows we will always take $\zeta=-1$ and write $\qq_{\,i}$ for $\qq_{\,i}^{(\zeta=-1)}$.

The $U(1)_r$ charge (or rather its double $d=2r$ in our conventions) plays the role of cohomological (homological) degree with respect to the action of $\qq_{\,1}$ ($\qq_{\,2}$). The (co)homology of these supercharges can then be identified with the vector space $\mathcal{V}$ of \emph{Schur operators} by taking harmonic representatives,\footnote{See Appendix~\ref{app:multiplets} for additional detail regarding the four-dimensional $\NN=2$ superconformal multiplets that contain Schur operators.}
\begin{equation}
    H^{\bullet}(\mathcal{H}^{(loc)},\qq_{\,1}) \cong H_{\bullet}(\mathcal{H}^{(loc)},\qq_{\,2})\cong \mathcal{V}^{\,\bullet}~.
\end{equation}
We recall that Schur operators are uniquely characterized by the fact that their quantum numbers obey the linear relations
\begin{equation}\label{eq:Schur_quantum_numbers}
    E = 2R + j_1 + j_2~,\qquad r = j_2 - j_1~.
\end{equation}
(In fact, the former condition implies the latter \cite{Beem:2013sza}.) Thus, the vector space underlying the associated vertex operator algebra has, in addition to cohomological grading, a further double grading which we will take to be by $R$-charge as well as the (chiral) conformal weight $h$, 
\begin{equation}
\begin{split}
    h = \tfrac12(E + j_1 + j_2) &= E - R~, \\
    &= R + j_1 + j_2~,
\end{split}
\end{equation}
and we write the corresponding weight decomposition as
\begin{equation}
    \mathcal{V}^{\,\bullet} = \bigoplus_{h,R\in\frac12\mathbb{N}}\mathcal{V}^{\,\bullet}_{h,R}~.
\end{equation}
To realize the vertex algebra structure on $\mathcal{V}^{\,\bullet}$, one utilizes \emph{twisted} translations in the $(z,\bar{z})$ plane to displace Schur operators from the origin,
\begin{equation}\label{eq:twistedtransl}
    \mathcal{O}(z) \coloneqq \left[e^{zL_{-1}+\bar{z}\widehat{L}_{-1}}\mathcal{O}(0)e^{-zL_{-1}-\bar{z}\widehat{L}_{-1}}\right]_{\qq_{\,i}}~,
\end{equation}
where $\widehat{L}_{-1} = \bar{L}_{-1} + \mathcal{R}_-$ is the ($R$-)twisted anti-holomorphic translation operator in the $(z,\bar z)$ plane (see Appendix~\ref{app:conventions} for precise definitions of these generators). In Euclidean signature, operators cannot be displaced in the transverse $(w,\bar{w})$ directions while remaining $\qq\,$-closed. (See, however, \cite{Argyres:2022npi} where more involved extensions in these directions or in null directions in Lorentzian signature are considered using supersymmetric descent.) The four-dimensional operator product expansion (OPE) of twisted-translated Schur operators is then meromorphic in $\qq\,$-(co)homology and defines a $\frac12\mathbb{N}$-graded vertex algebra structure in cohomology (with grading by the chiral weight $h$ and cohomological grading by twice $U(1)_r$ charge).

An important result of \cite{Beem:2013sza} is that the associated vertex algebra is canonically equipped with a conformal vector that extends the action of $\mathfrak{sl}(2)_z$ to a Virasoro action with central charge. 
\begin{equation}\label{eq:c2dc4d}
    c=-12c_{4d}~,
\end{equation}
where $c_{4d}$ is a Weyl anomaly coefficient of the four-dimensional parent theory (which equivalently determines the two-point function of the canonically normalized stress tensor). Thus, the vertex algebra structure in $\qq\,$ cohomology is actually a vertex \emph{operator} algebra structure. As the only Schur operator with $h=0$ is the identity operator, this VOA is of \emph{CFT type} (in fact, it follows from general considerations that it is of \emph{strong CFT type} \cite{moriwaki2020classification}, though this notion will not play a role in what follows).

\subsection{\label{subsec:R-filtration}\texorpdfstring{$\mathfrak{R}$}{R}-filtration and the associated vertex Poisson algebra}

While the vector space underlying the associated vertex operator algebra is canonically isomorphic to the vector subspace of Schur operators in the original SCFT, the $R$ grading on $\mathcal{V}^\bullet$ is not preserved in any simple fashion by the vertex algebra OPE. Indeed, the $R$ grading is not preserved under the twisted version of translation, which mixes four-dimensional operators with different $R$ charges.

However, the corresponding \emph{$R$-parity} of operators is determined just by conformal weight and cohomological degree. This is because for Schur operators we have
\begin{equation}
\begin{split}
    R &= h - j_1 - j_2~, \\ &= h - 2j_2 + \tfrac12 d~,\\ &\cong h + \tfrac12 d \mod\mathbb{Z}~.
\end{split}
\end{equation}
Thus an $R$-parity operator can be defined directly on the (conformally and cohomologically graded) associated vertex algebra without reference to $R$ grading (we denote this bare, $\tfrac12\mathbb{N}\times\mathbb{Z}$-graded vertex algebra by $\mathbb{V}^\bullet$); we define
\begin{equation}
\begin{split}
    s:\ & \mathbb{V}^\bullet\to\mathbb{V}^\bullet~,\\
    &a\mapsto (-1)^{2h + d} a~.
\end{split}
\end{equation}
The OPE is $s$-equivariant, so this constitutes a $\mathbb{Z}_2$ vertex algebra automorphism and gives rise to a decomposition $\mathbb{V}^\bullet=\mathbb{V}^\bullet_{\rm even}\oplus\mathbb{V}^\bullet_{\rm odd}$.\footnote{Note that in the case of vertex operator superalgebras there is a separate $\mathbb{Z}_2$ grading by Grassmann parity which need not be related to $R$-parity, though for the free vector multiplet they coincide.}

The rest of the $R$ grading on $\mathcal{V}$ can be used to define a filtration (the \emph{$\mathfrak{R}$-filtration}) on the vertex algebra $\mathbb{V}^{\bullet}$ that is better behaved than the $R$ grading itself,
\begin{equation}
    \mathfrak{R}_p\mathbb{V}_h^{\,\bullet} = 
    \begin{cases}
        \bigoplus_{R=0}^p \mathcal{V}^{\,\bullet}_{h,R}~,\qquad &p\in\mathbb{N}~,\\
        \bigoplus_{R=\frac12}^{p} \mathcal{V}^{\,\bullet}_{h,R}~,\qquad &p\in\frac12+\mathbb{N}~.
    \end{cases}
\end{equation}
Note that this is separately a filtration on $\mathbb{V}_{\rm even}$ and $\mathbb{V}_{\rm odd}$, as the $R$ grading violation in the vertex algebra OPE is by integers.

The $\mathfrak{R}$-filtration is increasing and exhaustive. It is also a \emph{good filtration} in the following sense.

\medskip

\begin{definition}[Generalized from Def. 4.1 of \cite{li2004vertex}]
    A \emph{half-integral good filtration} on a $\mathbb{Z}_2$-graded vertex algebra $V=V_{0}\oplus V_{\frac12}$ is an increasing, exhaustive filtration on $V_{0}$ and $V_{\frac12}$,
    \begin{equation}
    \cdots \subseteq F_{-1+\alpha}V_\alpha\subseteq F_{0+\alpha}V_\alpha\subseteq F_{1+\alpha}V_\alpha\subseteq \cdots~,\qquad \bigcup_{p\in\mathbb{Z}}^\infty F_{p+\alpha}V_{p} = V_\alpha~,\qquad \alpha=0,\tfrac12~,
    \end{equation}
    for which the identity vector $\Omega\in F_0V_0$ and such that for any $a\in F_{p+\alpha}V_\alpha$ and $b\in F_{q+\alpha}V_\beta$ one has\footnote{We recall here that the mathematical convention for the $n^{\rm th}$ product $a_{n}b$ of $a$ and $b$ is such that the normally ordered product is the $-1^{\rm st}$ product and the simple pole in the OPE is the $0^{\rm th}$ product. For the $\mathfrak{R}$-filtration, this condition says that the singular part of the OPE of two (twisted translated) Schur operators reduces $R$ grade by at least one, which follows from the absence of singularities in the OPEs of strict Schur operators.}
    \begin{equation}
    \begin{alignedat}{2}
        a_ib&\in F_{p+q+\alpha+\beta}V_{\alpha+\beta}~, \qquad&& i <0~,\\
        a_ib&\in F_{p+q+\alpha+\beta-1}V_{\alpha+\beta}~,\qquad &&i\geqslant 0~,
    \end{alignedat}
    \end{equation}
    where $V_{\frac12+\frac12}\equiv V_0$. A $\mathbb{Z}_2$ graded vertex algebra equipped with a half-integral good filtration is called a \emph{half-integer-filtered vertex algebra}.
\end{definition}

The filtration $\mathfrak{R}_\bullet$ is a half-integral good filtration (with respect to the $\mathbb{Z}_2$ grading by $R$-parity), and furthermore, $\mathfrak{R}_p\mathbb{V}=0$ for $p<0$ and $\mathfrak{R}_0\mathbb{V}=\mathbb{C}{\bf 1}$. Due to the $R$ nonequivariance of the vertex algebra OPE, it is this filtered vertex algebra structure that arises most naturally from the cohomological construction. We write $(\mathbb{V}^\bullet,\mathfrak{R}_\bullet)$ for the associated vertex algebra equipped with the $\mathfrak{R}$-filtration, without a given isomorphism with the graded space $\mathcal{V}^\bullet$ of Schur operators.

A useful result of \cite{li2004vertex}, which generalizes immediately to the half-integral setting, states that any such good filtration for a vertex algebra $V$ admits a standard description. Let $U\subset V$ be a (possibly infinite) vector subspace of $\mathcal{V}$ equipped with a grading $U=\coprod_{n\in\frac12\mathbb{N}}U_n$ such that $U$ (weakly) generates the vertex algebra. Define the filtration
\begin{equation}
    \mathfrak{W}^{U}_{p}V = {\rm span}\left\{u^{(1)}_{-1-n_1}\cdots u^{(k)}_{-1-n_k}{\bf 1} \quad\middle|\quad n_i\geqslant0~,\quad u^{(i)}\in U_{m_i} \quad \sum_{i=1}^k m_i\leqslant p\right\}~.
\end{equation}
If the following \emph{goodness condition} is satisfied,
\begin{equation}
    u_{i}v \in \mathfrak{W}^U_{p+q-1}V\quad {\rm for} \quad u\in U_{p}~,\quad v\in U_{q}~0,\quad i\geqslant0~,
\end{equation}
this will be a good filtration, and any good filtration with the aforementioned truncation conditions can be realized in this manner. We describe such a filtration as a \emph{weight-based filtration}, where the weights in question refer to the grades assigned to the weak generators in $U$.

For any filtered vertex algebra, by passing to the associated graded of the given filtration one recovers a Poisson vertex algebra \cite{li2004vertex}. In the context of four-dimensional SCFTs, the vertex Poisson algebra recovered from the $\mathfrak{R}$-filtration is precisely the local operator algebra of the holomorphic-topological twist of the same theory \cite{Kapustin:2006hi,Beem:String-Math-2017,Beem:Pollica,Beem:String-Math-2019,Oh:2019mcg}. In this sense, the VOA arising from four dimensions is a deformation quantization of the holomorphic-topological vertex Poisson algebra; this perspective is perhaps most natural when working in terms of a ($B$-type) $\Omega$-deformation \cite{Oh:2019bgz,Jeong:2019pzg}.

It is instructive to compare to several filtrations that frequently arise in the study of more general vertex algebras (\cf, \emph{e.g.}, \cite{arakawa2021arc}). Any vertex algebra $V$ admits a canonical \emph{decreasing} filtration known as \emph{Li's canonical filtration}. This is essentially a filtration by the number of derivatives appearing in a given expression of an element of $V$,
\begin{equation}
    \mathfrak{F}^pV = {\rm span}\left\{a^1_{-1-n_1}\cdots a^k_{-1-n_k}\Omega ~\middle|~ \sum n_k \geqslant p\right\}~.
\end{equation}
For a conformally graded vertex algebra (a $\frac12\mathbb{N}$-graded vertex algebra in our cases) there is an alternative \emph{conformal weight-based filtration} which is a good filtration in the sense of the above definition which tracks the conformal weights of the strong generators appearing in an expression for an element of $V$,
\begin{equation}\label{eq:weight_based_filtration_def}
    \mathfrak{W}_pV = {\rm span}\left\{a^1_{-1-n_1}\cdots a^k_{-1-n_k}\Omega ~\middle|~ \sum h_{a^i} \leqslant p\right\}~.
\end{equation}
This is a special case of a weight-based filtration where the graded subspace $U$ is taken to be the span of the strong generators and their gradings are given by their conformal weights. This is always a good filtration.

These two filtrations are complementary, in the sense that within a fixed conformal weight subspace one may be recovered from the other,
\begin{equation}
    \mathfrak{F}^pV\cap V_{\Delta} = \mathfrak{W}_{\Delta-p}\cap V_{\Delta}~.
\end{equation}
Correspondingly, their associated graded vertex Poisson algebras are isomorphic,
\begin{equation}
    \mathrm{gr}_{\mathfrak{F}^{\bullet}}V \cong \mathrm{gr}_{\mathfrak{W}_{\bullet}}V~.
\end{equation}

The $\mathfrak{R}$-filtration is a different example of a weight-based filtration, where rather than the defining subspace $U$ being graded by conformal weights, it is graded by $R$-charges. Because of the relation $R\leqslant h$, one has $\mathfrak{W}_p\mathcal V\subseteq \mathfrak{R}_p\mathcal V$, and one says that the $\mathfrak{R}$ filtration is \emph{coarser} than the conformal weight-based filtration. It is important (as we will review later in examples) that while the conformal weight-based filtration can always be defined in terms of a minimal set of strong generators, for the $\mathfrak{R}$-filtration one may need to take the defining subspace $U$ to be large. (For the purposes of definitions, one can always let $U=\mathcal{V}$  be the entire space of Schur operators, though it may be possible to use a smaller generating subspace.)

\medskip

\begin{remark}
    One can equivalently define a complementary decreasing filtration as an analogue of the Li canonical filtration as follows,
    \begin{equation}
        \Rdec^p\mathcal{V} = \bigoplus_{R\leqslant h-p} \mathcal{V}^{\,\bullet}_{h,R}~.
    \end{equation}
    This is manifestly complementary to the $\mathfrak{R}$-filtration in the same way that Li's canonical filtration is complementary to the conformal weight-based filtration, and gives rise to the same vertex Poisson algebra as its associated graded.
\end{remark}

It is an interesting open question to give an abstract characterization of the $\mathfrak{R}$-filtration that doesn't depend on extra information coming from four-dimensional physics. There is a proposal due to T.~Arakawa and A.~Moreau \cite{Arakawa_talks} for such a definition in certain restricted cases (from a physical point of view, for cases where the effective field theory on a generic point of the Higgs branch consists of only free hypermultiplets). We will not discuss that proposal further in this paper, but it would be interesting to study its interactions with the ideas appearing here.

\subsection{\label{subsec:conjugation}Conjugation and positivity}

Unitarity of a four-dimensional SCFT implies, at the level of the OPE of local operators, the existence of a $\mathbb{C}$-antilinear automorphism of the space of local operators arising from their realization as concrete operator(-valued distribution)s on a Hilbert space. Here we perform our analysis in ``planar quantization'', where the imaginary $z$ direction in the chiral algebra plane is taken to be Euclidean time, \emph{i.e.}, $z = x + i t_E$.\footnote{It is common in the vertex algebra literature, not to mention the conformal field theory literature, to work in radial quantization; a convenience of using planar quantization here is that we can naturally adopt conventions so that operators at the origin (where the space of Schur operators is naturally defined) remain at the origin under conjugation. Of course, the two constructions are related by a global conformal transformation and the positivity constraints described below can be equivalently formulated in radial quantization. This proves more useful for some other purposes \cite{BeemGarner:Hodge}.} Then taking the adjoint of a given local operator gives an involution on the space of local operators at the origin,
\begin{equation}
    \ast:\mathcal{H}^{(loc)}~\longrightarrow~\mathcal{H}^{(loc)}~.
\end{equation}
Of course, this involution does not preserve quantum numbers, and in fact does not preserve the subspace of Schur operators,
\begin{equation}
    \ast:\mathcal{H}^{(loc)}_{E,j_1,j_2,R,r}~\longrightarrow~ \mathcal{H}^{(loc)}_{E,j_2,j_1,-R,-r}~.
\end{equation}
The failure of \eqref{eq:Schur_quantum_numbers} on the right hand side boils down to the sign of the $R$-charge being wrong. This can be rectified by defining an improved conjugation automorphism,\footnote{The precise choice of rotation-by-$\pi$ in $R$-symmetry space is a convention. We have chosen a convention so that the highest weight state $\mathcal{O}_{h.w.}$ of a spin-$R_\mathcal{O}$ $\mathfrak{su}(2)_R$ multiplet will be sent to a positive real multiple of $\mathcal{R}_-^{2R_\mathcal{O}}\mathcal{O}_{h.w.}$.}
\begin{equation}\label{eq:conjdef}
\begin{split}
    \Rho:~&\mathcal{H}^{(loc)}~\longrightarrow~\mathcal{H}^{(loc)}~,\\
    &\mathcal{O}(0)~\longmapsto~ e^{-\pi i \mathcal{R}_2}\mathcal{O}^\ast(0) e^{\pi i \mathcal{R}_2}~.
\end{split}
\end{equation}
The extra rotation by $\pi$ in $R$-symmetry space negates $R$-charge again, so in particular we have
\begin{equation}
    \Rho:\mathcal{V}^{\,\bullet}~\longrightarrow~\mathcal{V}^{-\bullet}~,
\end{equation}
where cohomological degree is reversed. In the presence of half-integral $R$-charges $\Rho$ will not be an involution, but rather an order-four anti-linear automorphism that squares to the $R$-parity/degree operator,
\begin{equation}
    \Rho\circ\Rho = (-1)^{2R} \eqqcolon s~.
\end{equation}
It will be an involution on $\mathbb{V}_{\rm even}$ and square to minus the identity on $\mathbb{V}_{\rm odd}$.

\subsubsection{Vertex algebra properties of conjugation}

A crucial feature of the automorphism $\Rho$ is that it interacts nicely with the vertex algebra structure on $\qq$ cohomology. First, we note that $\Rho$ intertwines the action of the key supercharges in the $\mathfrak{sl}(2|4)$ superconformal algebra according to
\begin{equation}
\begin{split}
    \Rho\circ\mathcal{Q}^{1}_{-} &~=~ \widetilde{\mathcal{Q}}_{2\dot{-}}\circ\Rho~,\\
    \Rho\circ\widetilde{\mathcal{S}}^{2\dot{-}} &~=~ -\mathcal{S}_{1}^{-}\circ\Rho~.
\end{split}
\end{equation}
Thus, the action of $\Rho$ intertwines the action of $\qq_{\,1}$ and $\qq_{\,2}$ and induces a map on cohomology,
\begin{equation}
    \rho:\ H^\bullet(\mathcal{H}^{(loc)},\qq_{\,1})~\longrightarrow~H^{-\bullet}(\mathcal{H}^{(loc)},\qq_{\,2})~.
\end{equation}
This action can be extended to accommodate twisted translation at the expense of replacing the complex translation by $z$ by a translation by the complex conjugate $z^\ast$,
\begin{equation}
    \rho\circ \left(\left[\mathcal{O}\right]_{\qq_{\,1}}\!\!(z)\right) = \left[\Rho(\mathcal{O})\right]_{\qq_{\,2}}\!\!(z^\ast)~.
\end{equation}
As a consequence, $\rho$ actually defines an anti-linear automorphism of the vertex algebra $\mathbb{V}$. 

An immensely useful consequence of the fact that $\rho$ is a vertex algebra automorphism is that the conjugate of the normally ordered product of two operators is determined by the conjugate of those operators. In particular (including signs relevant for fermionic operators), we have
\begin{equation}\label{eq:conjugation_for_composites}
    \rho\circ(ab) = (-1)^{p(a)p(b)}(\rho(a)\rho(b))~,
\end{equation}
where $p$ is the Grassmann parity operator. Consequently, for strongly finitely generated vertex algebras, the action of $\rho$ need be specified only for the strong generators.

\subsubsection{Sesquilinear, Hermitian, and positive forms}

Given the conjugation $\rho$, we define a sesquilinear form $(,):\mathbb{V}_h\times\mathbb{V}_h\to\mathbb{C}$ by taking the coefficient of two-point functions \footnote{See, \emph{e.g.}, \cite{dong2014unitary} for the formalized structure of unitarity for vertex operator algebras, though formulated in radial quantization; here \emph{we restrict the pairing to operators of the same conformal weight by hand}.},
\begin{equation}\label{eq:sesquilinear_form_def}
    (a,b)\coloneqq \left\langle a(\tfrac{i}{2}) (\rho\circ b)(-\tfrac{i}{2})\right\rangle~,\qquad a,b\in\mathcal{V}_h~,
\end{equation}
where the insertion points are chosen to be related by reflection through the Euclidean time $t_E=0$ hypersurface. This form is not necessarily Hermitian; indeed we have
\begin{equation}
\begin{split}
    \overline{(a,b)} &= \langle \rho\big(a(\tfrac{i}{2})\,(\rho\circ b)(-\tfrac{i}{2})\big)\rangle~,\\[2pt]
    &= \langle (\rho^2\circ b)(\tfrac{i}{2})\,(\rho\circ a)(-\tfrac{i}{2})\rangle~,\\[2pt]
    &= \langle (s\circ b)(\tfrac{i}{2})\,(\rho\circ a)(-\tfrac{i}{2})\rangle~,\\[2pt]
    &= (-1)^{2R_b}(b,a)~.
\end{split}
\end{equation}
So for $a$, $b$ of even $R$ parity, this is an Hermitian form, but for $a,b$ of odd $R$ parity it is anti-Hermitian.%
    \footnote{This can be understood quite explicitly. By translation invariance of the vertex algebra, we can equally evaluate these pairings with insertions at $0$ and $-i$. Then the difference between this pairing and the usual Hilbert space pairing of $a$ and $b^\ast$ is a factor of $(i)^{2R_b}$ that comes from the twisted-translation definition of vertex algebra operators away from the origin. For integer $R_b$ this has no effect on the Hermiticity of the pairing, but for half-integer $R_b$ an extra minus sign arises when taking complex conjugates.} 
Even for the Hermitian case, though, this form need not be positive definite as given. 

Both Hermiticity and positive-definiteness can be recovered by introducing a further $\mathbb{Z}_4$ grading on $\mathbb{V}$ which amounts to knowing the $R$-\emph{charges} $\mathrm{mod}~4$,
\begin{equation}
\begin{split}
    \sigma:\,&\mathcal{V}\to\mathcal{V}~,\\
    &a\mapsto (i)^{2R_a} a ~,\qquad a \in \mathcal{V}^{\bullet}_{h_a,R_a}~,
\end{split}
\end{equation}
so we are using our knowledge of $R$ charges to define a square root of the $R$ parity, $\sigma^2 = s$. We can then define an improved sesquilinear form on $\mathcal{V}$ according to
\begin{equation}\label{eq:hermitianform}
    \bm{(}a\mathbf{\,,\,}b\bm{)} \coloneqq \langle a(\tfrac{i}{2})\,(\sigma\circ\rho\circ b))(-\tfrac{i}{2})\rangle~,
\end{equation}
and verify that the resulting bilinear form is in fact Hermitian,
\begin{equation}
    \overline{\bm{(}a\mathbf{\,,\,}b\bm{)}} = \bm{(}b\mathbf{\,,\,}a\bm{)}~.
\end{equation}
Note that even for $a,b\in\mathbb{V}_{\rm even}$, this form may differ by an overall sign from the naive sesquilinear pairing.

Crucially, this form is in fact positive definite. This is simplest to observe by using translation invariance to rewrite the above as
\begin{equation}\label{eq:translated_hermitian_form}
    \bm{(}a\mathbf{\,,\,}b\bm{)} = \langle a(i)\,(\sigma\circ\rho\circ b))(0)\rangle~.
\end{equation}
Now suppose $a\in\mathcal{V}$ corresponds to a Schur operator $A_{h.w.}$, which is in an $\mathfrak{su}(2)_R$ highest weight state. Then we have
\begin{equation}
\begin{split}\label{eq:positivity_check}
    \bm{(}a\mathbf{\,,\,}a\bm{)} &= \langle a(i)\,(\sigma\circ\rho\circ a))(0)\rangle~.\\
    & \propto (-i)^{2R_A}(i)^{2R_A} \langle A_{l.w.}(i)(A_{l.w.})^\ast(0)\rangle\\
    & \propto \langle A_{l.w.}(i)(A_{l.w.})^\ast(0)\rangle~,
\end{split}
\end{equation}
where all proportionalities are with positive real constant. Thus, we have a result that is a positive real multiple of a four-dimensional two-point function that is positive definite by reflection positivity.

\subsubsection{\label{subsec:practical_formula}A practical formula}

In concrete applications it is useful to formulate the positivity properties of the above Hermitian form as a simple rule for the coefficients of the vertex algebra two-point functions of operators and their conjugates,
\begin{equation}
    \langle \mathcal{O}(z)(\rho\circ\mathcal{O})(0)\rangle \eqqcolon \frac{\kappa_\mathcal{O}}{z^{2h_\mathcal{O}}}~.
\end{equation}
Positive definiteness of the above inner product can then be expressed as the requirement,
\begin{equation}\label{eq:practical_rule}
    \kappa_\mathcal{O}\propto (i)^{2h_\mathcal{O}-2R_\mathcal{O}} = (i)^{2(j_1+j_2)}~,
\end{equation}
where as above, proportionality means up to a real positive multiple and we have used the identity \eqref{eq:Schur_quantum_numbers} for the quantum numbers of Schur operators in the latter equality. For bosonic operators, this becomes a simple \emph{sign rule} for two-point function coefficients that we will use extensively later in this paper,
\begin{equation}\label{eq:bosonic_sign_rule}
    \kappa_\mathcal{O}\propto (-1)^{h_\mathcal{O}-R_\mathcal{O}}~,\qquad \text{($\mathcal{O}$ bosonic)}~.
\end{equation}

\subsubsection{\label{subsubsec:Gram-Schmidt}Nondegeneracy, recovering the $R$ grading, and graded unitarity}

The definition of the positive definite inner product $\bm{(}-\mathbf{\,,\,}-\bm{)}$ makes use of the $R$ grading (mod 4). However, the entire structure we have defined to this point can be formulated entirely at the level of the $\mathfrak{R}$-filtered vertex algebra (with conjugation). In particular, the sesquilinear form \eqref{eq:sesquilinear_form_def} is well defined in this context, and although it is not yet positive definite, it is \emph{required} by four-dimensional unitarity to be nondegenerate. Furthermore, the $\mathfrak{R}$-filtration must be nondegenerate in the sense that \eqref{eq:sesquilinear_form_def} restricts to a nondegenerate form on each component of the filtration $\mathfrak{R}_p$. If this is the case, then by a generalization of the Gram--Schmidt process the actual $R$ grading can be recovered, and so the positive definite form \eqref{eq:hermitianform} can be defined.

Thus we finally arrive at the definition of what we will call a {\it graded unitary} vertex algebra:

\medskip

\begin{definition}[Graded unitary vertex algebra]
    A \emph{graded unitary} vertex algebra is a triple $(\mathbb{V}^\bullet,\mathfrak{R}_\bullet,\rho)$, where:
    \begin{itemize}
        \item[(i)] $\mathbb{V}^\bullet=\bigoplus_{h}\mathbb{V}^\bullet_h$ is a $\frac12\mathbb{N}$-graded vertex algebra with integer cohomological grading $d$ and corresponding $R$-parity operation $s=(-1)^{h+\frac12 d}$,
        \item[(ii)] $\rho$ is an anti-linear vertex-algebra automorphism obeying $\rho^2 = s$.
        \item[(iii)] $(,):\mathbb{V}\times\mathbb{V}\to\mathbb{C}$ is the sesquilinear form defined (for each conformal weight space) as in \eqref{eq:sesquilinear_form_def} using $\rho$,
        \item[(iv)] $\mathfrak{R}_\bullet$ is a half-integral good filtration on $\mathbb{V}$ with respect to which the sesquilinear form is nondegenerate,
        \item[(v)] $\mathbb{V}_h^\bullet = \bigoplus_{R=0}^h \mathbb{V}^\bullet_{h,R}$ is the corresponding orthogonal decomposition of a given conformal weight space into orthogonal graded subspaces resulting from applying the Gram--Schmidt process using $(,)$,
        \item[(vi)] $\bm{(}\mathbf{\,,\,}\bm{)}:\mathbb{V}\times\mathbb{V}\to\mathbb{C}$ is the Hermitian form defined as in \eqref{eq:hermitianform} and is \emph{positive definite}.
    \end{itemize}
\end{definition}

Beyond this definition, there are additional features of the $\mathfrak{R}$-filtration that follow from the idiosyncrasies of short multiplets in four-dimensional $\mathcal{N}=2$ SCFTs. In particular, only certain types of operators are allowed to appear at low values of $R$ charge, and $R$ charge is bounded above by conformal weight (and by conformal weight minus $U(1)_r$ charge for states with nonzero cohomological degree). We translate these requirements into a further set of restrictions on the $\mathfrak{R}$-filtration.

\medskip

\begin{definition}[$\mathfrak{R}$-filtration of four-dimensional type]
    We say that a half-integral good filtration on a vertex algebra $\mathbb{V}$ is of four-dimensional type if it satisfies the following requirements:
    \begin{itemize}
        \item[(i)] $\mathfrak{R}_p\mathbb{V}=0~,\qquad p<0$;
        \item[(i)] $\mathfrak{R}_0\mathbb{V}=\mathbb{C}\Omega$;
        \item[(ii)] $\mathfrak{R}_{\frac12}\mathbb{V}$ is the span of the strong generators of any symplectic bosons and/or symplectic fermions subalgebras of $\mathbb{V}$;
        \item[(iii)] $\mathfrak{R}_1\mathbb{V}$ includes the span of the conformal vector if $\mathbb{V}$ is a vertex operator algebra, as well as any affine currents and any supercurrents (if the Virasoro algebra is enhanced to the $\mathcal{N}=2$ or small $\mathcal{N}=4$ super-Virasoro algebra);
        \item[(iv)] $\mathfrak{R}_p\mathbb{V}^d_h=0~,\qquad p+\tfrac12 |d| > h$.
    \end{itemize}
\end{definition}

In general, if a four-dimensional theory has free fields (corresponding to symplectic boson or fermion subalgebras of the associated vertex algebra) or discrete quotients thereof, then there may be additional states in $\mathfrak{R}_1\mathbb{V}$. These states correspond to higher-spin conserved current multiplets in four dimensions and will therefore be absent in intrinsically interacting theories. Thus, we further define interacting $\mathfrak{R}$-filtrations,

\medskip

\begin{definition}[Interacting $\mathfrak{R}$-filtration]
    We say that an $\mathfrak{R}$-filtration of four-dimensional type is an \emph{interacting} $\mathfrak{R}$-filtration if $\mathfrak{R}_{\frac12}\mathbb{V}=0$ (so no free fields) and $\mathfrak{R}_1\mathbb{V}$ is exactly the span of the vacuum vector, the conformal vector if $\mathbb{V}$ is a vertex operator algebra, any affine currents, and any supercurrents (if the Virasoro algebra is enhanced to the $\mathcal{N}=2$ or small $\mathcal{N}=4$ super-Virasoro algebra). 
\end{definition}

\subsection{\label{subsec:free_fields}Simple examples}

The above structures can be illustrated in the simple case of free field theories. Here we detail the cases of the free hypermultiplet (corresponding to a symplectic boson VOA) and the free vector multiplet (corresponding to the symplectic fermion VOA).

\subsubsection{Free hypermultiplet and symplectic bosons}

Recall that for the free hypermultiplet, the basic Schur operators are the complex scalars $Q$ and $\tilde Q$ that sit in $\mathfrak{sl}(2)_R$ doublets as follows (up to a choice of conventions),
\begin{equation}
    Q^I = \begin{pmatrix}Q\\-\tilde{Q}^\ast\end{pmatrix}~,\qquad
    \tilde{Q}^I = \begin{pmatrix}\tilde{Q}\\Q^\ast\end{pmatrix}~.
\end{equation}
The corresponding twisted-translated Schur operators $q$ and $\tilde{q}$ then obey the symplectic boson OPE,
\begin{equation}
    q(z)\tilde{q}(w) = \frac{-1}{z-w}+ \reg~.
\end{equation}
With our conventions from above, the conjugation automorphism acts according to
\begin{equation}
    \rho(q) = - \tilde{q}~,\qquad \rho(\tilde q) = q~.
\end{equation}
The positive-definite Hermitian form, restricted to the space spanned by $q$ and $\tilde{q}$, is then given by
\begin{equation}
\begin{split}
    \begin{pmatrix}
    \bm{(}q,q\bm{)} & \bm{(}q,\tilde{q}\bm{)}\\
    \bm{(}\tilde{q},q\bm{)} & \bm{(}\tilde{q},\tilde{q}\bm{)}
    \end{pmatrix} &= 
    \begin{pmatrix}
    (i)\langle q(\tfrac{i}{2})(-\tilde{q})(-\tfrac{i}{2})\rangle & 0\\
    0 & (i)\langle \tilde{q}(\tfrac{i}{2})q(-\tfrac{i}{2})\rangle
    \end{pmatrix}\\
    & = \mathrm{Id}_{2\times 2}~.
\end{split}
\end{equation}
Alternatively, using the simplified formula from above we have
\begin{equation}
    \kappa_q = \kappa_{\tilde q} = 1 \propto i^0~,
\end{equation}
which verifies compatibility with unitarity in this case.

\subsubsection{Free vector multiplet}

The case of the free vector is very similar, though factors of $i$ require careful attention. Here there are two basic Schur operators, the (positive-helicity) Weyl fermions $\lambda_+^1$ and $\tilde{\lambda}_{\dot{+}}^1$ that live in multiplets $\lambda_\alpha^I$ and $\tilde{\lambda}_{\dot{\beta}}^J$. The corresponding twisted-translated vertex operators are denoted $\eta_+$ and $\eta_-$, respectively, and obey the symplectic fermion OPE \cite{Beem:2013sza},
\begin{equation}\label{eq:sf_OPE}
    \eta_+(z)\eta_-(w)=\frac{-1}{(z-w)^2}+\reg ~.
\end{equation}
It turns out that the normalization for the gauginos that leads to precisely this OPE (taken from \cite{Beem:2013sza}) is somewhat unconventional from a four-dimensional perspective, and indeed the action of conjugation is given by\footnote{More conventional would be to adopt normalizations where there is a factor of $-i$ in the OPE coefficient in \eqref{eq:sf_OPE} and then one would have $\rho(\eta_{\pm})=\mp\eta_{\mp}$, in closer parallel with the symplectic boson case.}
\begin{equation}
    \rho(\eta_+) = -i \eta_-~,\qquad \rho(\eta_-) = i \eta_+~.
\end{equation}
With this action in place, direct computation again yields a positive definite form on the two-dimensional space spanned by $\eta_\pm$,
\begin{equation}
\begin{split}
    \begin{pmatrix}
    \bm{(}\eta_+,\eta_+\bm{)} & \bm{(}\eta_+,\eta_-\bm{)}\\
    \bm{(}\eta_-,\eta_+\bm{)} & \bm{(}\eta_-,\eta_-\bm{)}
    \end{pmatrix} &= 
    \begin{pmatrix}
    (i)\langle\eta_+(\tfrac{i}{2})(-i\eta_-)(-\tfrac{i}{2})\rangle & 0\\
    0 & (i)\langle \eta_-(\tfrac{i}{2})(i\eta_+)(-\tfrac{i}{2})\rangle
    \end{pmatrix}\\
    & = \mathrm{Id}_{2\times 2}~.
\end{split}
\end{equation}
Again, in terms of the simplified formula we have
\begin{equation}
    \kappa_{\eta_+} = \kappa_{\eta_-} = i \propto (i)^{2-1}~,
\end{equation}
verifying compatibility with unitarity in this case.

\section{\label{sec:virasoro}Graded unitarity for Virasoro VOAs}

We now come to the question of graded unitarity for Virasoro vertex operator algebras. Throughout this section, we will be considering a hypothetical four-dimensional SCFT $\mathcal{T}_c$ whose associated VOA is isomorphic to precisely the Virasoro VOA at central charge $c$, with strong generator $T$ satisfying the usual OPE,
\begin{equation}
    T(z) T(0) \sim \frac{c/2}{z^4} + \frac{2 T(0)}{z^2}+ \frac{\partial T(0)}{z}~.
\end{equation}
The Virasoro central charge $c$ would be related to the four-dimensional central charge $c_{4d}$ as in \eqref{eq:c2dc4d}, and by nondegeneracy of two-point functions this should be the simple quotient vertex algebra,
\begin{equation}
\mathbb{V}[\mathcal{T}_c] \cong \text{Vir}_c\equiv L(c,0)~.
\end{equation}

There exists a single family of four-dimensional SCFTs that are generally understood to give rise to precisely Virasoro VOAs as their associated vertex algebras. These are the Argyres--Douglas theories of the type $(A_1,A_{q-3})$ with $q\geqslant5$ odd, whose associated vertex algebras coincide with those of the $(2,q)$ (nonunitary) Virasoro minimal models (with central charge $c=c_{2,q}\coloneqq13-\frac{12}{q}-3q$). It is presently unknown whether any other Virasoro VOAs arise as associated VOAs of four-dimensional theories.

As the full Virasoro vertex algebra is strongly generated by the stress tensor, the action of $\rho$ on the full VOA is determined by its action on $T$. This can be recovered from \eqref{eq:conjdef} using the four-dimensional origin of the stress tensor as the $\mathfrak{su}(2)_R$ current \cite{Beem:2013sza}, or alternatively by demanding that it is an (anti-)automorphism of the VOA. Either way, we find that conjugation acts trivially on the stress tensor,
\begin{equation}\label{eq:conjT}
    \rho\left(T\right) = T~,
\end{equation}
which implies that the conjugation automorphism acts according to $\rho(\mathcal{O}(z)) = \mathcal{O}(\bar z)$ for general operators (written in terms of $T$ with real coefficients) in the VOA.

\subsection{\label{subsec:vir_unitarity_ex}Constraints from graded unitarity at low level}

To begin, let us consider the constraints from graded unitary for low-dimension quasi-primary operators in the Virasoro algebra. The stress tensor itself is the only operator with conformal weight $h=2$, and for this we compute (recalling that the stress tensor has unit $R$-charge),
\begin{equation}\label{eq:cLeq0}
    \bm{(}T,T\bm{)} = (-1) \langle T(\tfrac{\ii}{2})T(-\tfrac{\ii}{2})\rangle =- \frac{c}{2} \geqslant 0\qquad   \Longleftrightarrow \qquad c\leqslant0~.
\end{equation}
By \eqref{eq:c2dc4d} this is precisely the four-dimensional unitary requirement that $c_{4d} \geqslant 0$.

At dimension three, the only operator is $T'$ which is a descendant and whose positivity is guaranteed by that of the quasi-primary $T$. Proceeding to the next nontrivial level, there is a single dimension-four quasi-primary\footnote{From here onward we adopt the convention that a product of local operators $\OO_1 \OO_2\ldots \OO_n$ means a nested normal ordered product $(\OO_1( \OO_2 \ldots (\OO_{n-1} \OO_n)))$, so $T^2 = (T T)$ in \eqref{eq:level-four-quasi}.}
\begin{equation}\label{eq:level-four-quasi}
    T^2_{(4)} \coloneqq T^2-\frac{3}{10}T''~.
\end{equation}
On general grounds, this could be some linear combination of Schur operators with $R$-charge one and two (the four-dimensional OPE selection rules for the self-OPE of two $\mathfrak{su}(2)_R$ currents are summarized in Appendix~\ref{app:multiplets}). Such a dimension-four quasi-primary with $R$-charge either one or two could arise from a Schur operator in either a $\hat{\CC}_{0(1,1)}$ multiplet or a $\hat{\CC}_{1(\frac12,\frac12)}$ multiplet, respectively (see Table~\ref{tab:schurmult}). The former contains conserved currents of spin greater than two, which are absent in interacting theories \cite{Maldacena:2011jn,Alba:2013yda}. We conclude that, in an interacting theory, $T^2_{(4)}$ must arise from a $\hat{\CC}_{1(\frac12,\frac12)}$ multiplet and has exactly $R=2$. Said differently, assuming $\mathfrak{R}_\bullet$ to be an interacting $\mathfrak{R}$-filtration on $L(c,0)$, we must assign $R$ charge two to $T^2_{(4)}$.

Computing the relevant two-point function, we now have
\begin{equation} 
    \bm{(}T^2_{(4)},T^2_{(4)}\bm{)} =  (-1)^2\langle T^2_{(4)}(\tfrac{\ii}{2}) T^2_{(4)}(-\tfrac{\ii}{2}) \rangle = \tfrac{1}{10} c (5 c+22) \geqslant 0~,
\end{equation}
which, given the negativity of $c$, amounts precisely to the unitarity bound of \cite{Liendo:2015ofa},
\begin{equation}\label{eq:clessc25}
    c \leqslant - \frac{22}{5} = c_{2,5}~.
\end{equation}
We note that the argument above did not require that the Virasoro algebra be the full vertex algebra in question, only that the corresponding SCFT be interacting (and not include any additional free fields). Indeed this bound was proved in \cite{Liendo:2015ofa} assuming only the absence of higher-spin conserved currents.

Moving to the next nontrivial level, there is a two-dimensional space of quasi-primaries at dimension six. A natural basis for these is given as follows.
\begin{equation}\label{eq:dim6Virquasi}
\begin{split}
        T_{(6)}^2& \coloneqq  T'' T-\tfrac{5}{4}T'^2+\tfrac{5}{4\times42}T''''~,\\ 
        T_{(6)}^3& \coloneqq  T^3+\tfrac{93}{29+70c}T'^2-\tfrac{3(67+42c)}{2(29+70c)}T'' T-\tfrac{13+10c}{4(29+70c)}T''''~.
\end{split}
\end{equation}
Here $T^2_{(6)}$ is the unique quasi-primary that can be expressed without using the $T^3$ operator, and $T^3_{(6)}$ is chosen orthogonal to $T^2_{(6)}$.\footnote{\label{footnote:degenerate_case}The poles in the expression for $T^3_{(6)}$ at $c=-\frac{29}{70}$ do not correspond to any null state, but rather to a problem in the choice of basis. This value of $c$ is ruled out by eq. \eqref{eq:clessc25} so the basis chosen is correct for our purposes. However, at $c=-\frac{29}{70}$ it is not possible to have an orthogonal basis of quasi-primaries such that one contains $T^3$ and the other does not, which is reflected in the appearance of the poles. There is a basis of orthogonal quasi-primaries but both operators contain $T^3$, and such a basis is incompatible with our assumed filtration.}

By the properties of the $\mathfrak{R}$-filtration, $T^2_{(6)}$ will have $R\leqslant 2$, and therefore must have $R=2$ given an interacting filtration. On the other hand, $T^3_{(6)}$ could in principle have either $R=2$ or $R=3$.\footnote{Here we are using the assumption that the VOA in question is precisely the Virasoro VOA; therefore this state must be a homogeneous vector with respect to the $R$-grading, rather than a possible mixture.} (Indeed, according to four-dimensional selection rules, this state could sit in a $\hat{\CC}_{1(\frac32,\frac32)}$ or $\hat{\CC}_{2(1,1)}$ multiplet.) The form \eqref{eq:hermitianform} evaluated on each of these two quasi-primaries gives
\begin{equation}\label{eq:level_six_quasiprimary_norms}
\begin{split}
    \bm{(}T_{(6)}^2,T_{(6)}^2\bm{)} &= \frac{9 c (70 c+29)}{28}~,\\
    \bm{(}T_{(6)}^3,T_{(6)}^3\bm{)} &= \pm\frac{3 c (2 c-1) (5 c+22) (7 c+68)}{4 (70 c+29)}~.
\end{split}
\end{equation}
The first quantity in \eqref{eq:level_six_quasiprimary_norms} must be positive by graded unitarity. This is already a consequence of \eqref{eq:clessc25}. In the second expression, the plus sign corresponds to the assignment $R[T^3_{(6)}]=2$ and the minus sign to $R[T^3_{(6)}]=3$. Combined with \eqref{eq:clessc25} we then have the following relationship between $R$ charge assignments and allowed central charge values: 
\begin{equation}\label{eq:startofinductionVir}
\begin{split}
    c = c_{2,5}~~\text{or}~~c = c_{2,7}~,\qquad & T^3_{(6)}=0~,\\
    c_{2,7}<c<c_{2,5}~,\qquad & R[T^3_{(6)}]=2~,\\
    c\leqslant c_{2,7}~,\qquad & R[T^3_{(6)}]=3~.
\end{split}
\end{equation}

\subsection{\label{subsec:virasoro_R_filt}A candidate \texorpdfstring{$\mathfrak{R}$}{R}-filtration}

The preceding analysis shows that the assignment of $R$ charges (by way of specifying an $\mathfrak{R}$-filtration) is crucial to determining constraints on the allowed Virasoro central charge. An ambitious program would be to constrain the possible $\mathfrak{R}$-filtrations \emph{ab initio} in conjunction with the requirements of graded unitarity. This appears to be a challenging undertaking, and even performing an analysis level-by-level quickly becomes very involved and suggests that additional ideas are needed.

Nevertheless, there exists a particular, natural filtration on the Virasoro vertex algebra that has a good case for being the true $\mathfrak{R}$-filtration inherited from four dimensions. This is a \emph{weight-based filtration} of the type first defined in \cite{li2004vertex} and reviewed in \ref{subsec:R-filtration}, though notably not the \emph{conformal weight filtration}. We define the filtration as follows,\footnote{Note that the Virasoro modes $L_n$ are defined as in \eqref{eq:modeexp}.}
\begin{equation}
    \mathcal{G}_p\mathcal{V} = \text{span}\left\{L_{-2-n_1}\cdots L_{-2-n_k}\Omega~|~n_i\geqslant0~,~k\leqslant p\right\}~.
\end{equation}
Informally, this counts the maximum number of copies of $T$ (or its derivatives) in an expression for an operator as the sum of normally ordered products of derivatives of $T$, where normal ordering ambiguities can only introduce terms with fewer copies of $T$ and null states allow for an operator to be placed in a lower $p$ filtered component of $\mathbb{V}$. It is immediate to check that this is a good filtration on $\mathbb{V}=L(c,0)$. What's more, this is the \emph{finest possible} good filtration compatible with the assignment of $R$-charge one to the stress tensor.\footnote{An increasing filtration $F_\bullet V$ is finer than a second increasing filtration $G_\bullet V$ if $F_pV\subseteq G_pV$ for all $p$.}

An interesting feature of this filtration is that it gives rise to a particularly simple Poisson vertex algebra when passing to the associated graded algebra. In particular, the resulting Poisson vertex algebra is generated entirely by the (commutative version of the) stress tensor. One can see fairly easily that for any alternative good $\mathfrak{R}$-filtration that one might propose for the Virasoro VOA, the associated graded would necessarily have additional generating fields. Thus, our proposed filtration is \emph{unique subject to the assumption that the associated graded vertex Poisson algebra be generated by the weight-one stress tensor}.

We note that this is precisely this filtration that was (implicitly) proposed for the $(2,2n+3)$ Virasoro VOAs (arising as the associated VOAs for the $(A_1,A_{2n})$ Argyres--Douglas SCFTs) in \cite{Song:2016yfd}, where it was checked explicitly up to moderate orders of the $q,t$-expansion by comparing with a conjectured closed form expression for the Macdonald index.\footnote{See also \cite{Foda:2019guo} for an alternate perspective on recovering the Macdonald index from Virasoro VOAs.} 

In what follows, \emph{we will take as a standing assumption that the $\mathfrak{R}$-filtration arising from our hypothetical four-dimensional SCFT is exactly the weight-based filtration given above}, \emph{i.e.}, $\mathfrak{R}_\bullet \stackrel{!}{=} \mathcal{G}_\bullet$. (We note here that this filtration is not always nondegenerate, and so by this assumption we are also excluding certain discrete values of the central charge. For example, at $c=-29/70$ the filtration becomes degenerate as described in Footnote \ref{footnote:degenerate_case}.)

With this assumption in place, an $R$-grading is uniquely determined by applying the Gram--Schmidt process with respect to the inner product \eqref{eq:sesquilinear_form_def} to the space of operators at a given conformal weight. Namely, we will have
\begin{equation}
    \mathcal{V}_{R,h}\coloneqq \left(\mathfrak{R}_R\mathcal{V}_h\right)\,\bigcap\,\left(\mathfrak{R}_{R-1}\mathcal{V}_h\right)^\perp
\end{equation}
In the instance of the $h=6$ subspace considered at the end of the previous subsection, this procedure designates $T^3_{(6)}$ as having $R=3$, and so rules out (with this choice of $\mathfrak{R}$-filtration) Virasoro central charges in the range $\big(c^{}_{2,7}\,,\,c^{}_{2,5}\big)$.

\subsection{\label{subsec:vir_unitarity_kac}Only \texorpdfstring{$(2,q)$}{(2,q)} minimal models}

Proceeding incrementally to higher levels reveals a pattern similar to that seen thus far. Namely, up to level $h=12$ one finds that with the given $\mathfrak{R}$-filtration, the level-$L$ graded unitarity constraints restrict to either $c=c_{2,q}$ with $q \in \{5,\ldots, L-1\}$, or $c \leqslant c_{2,L+1}$. At high levels, performing such an analysis explicitly becomes cumbersome. 

In this subsection we show that, still with the presumed $\mathfrak{R}$-filtration, only the $(2,q)$ minimal models are consistent with graded unitarity. To do this, we deduce from the assumption of graded unitarity the sign of the Kac determinant at a given level. We then show that for any central charge not equal to one of those of the $(2,q)$ minimal models, there is a level where the actual sign of the Kac determinant does not match that required by graded unitarity. Namely, we demonstrate that for $L\in2\mathbb{Z}_{\geqslant2}$, if $c<c_{2,L-1}$, then graded unitarity at level $L$ implies that $c\leqslant c_{2,L+1}$. In conjunction with \eqref{eq:cLeq0} this rules out all central charge values except for~$\{c_{2,L+1}\}$.

Recall that the Kac determinant at level $L$, denoted $\mathrm{det}M^{(L)}$, computes the determinant of the Shapovalov form for states at level $L$. The Shapovalov form can be understood as defining the pairing of states in the Virasoro algebra in radial quantization, where the adjoint of a Virasoro mode is given by $L_{n}^\dagrad = L_{-n}$. Of course, this pairing differs from the one used to define \eqref{eq:hermitianform}, and therefore does need not be positive for a graded unitary Virasoro VOA.

Four-dimensional primary operators give rise to $\mathfrak{sl}(2)$ primary operators in the VOA, and accordingly we organize Virasoro descendants of the vacuum into quasi-primaries and their $\mathfrak{sl}(2)$ descendants. A quasi-primary $\OO$ of dimension $L$ will have two-point function of the form
\begin{equation}
    \langle \OO(z) \OO(0) \rangle = \frac{\kappa_{\mathcal{O}}}{z^{2L}}~.
\end{equation}
The norm of the state created by $\OO(0)$ in radial quantization will be (see Appendix~\ref{app:Dagger})
\begin{equation}
    \langle\OO| \OO\rangle = (-1)^L \lim\limits_{z\to \infty} z^{2L} \langle \OO(z) \OO(0) \rangle =(-1)^L \kappa_{\mathcal{O}}~,
\end{equation}
The $\mathfrak{sl}(2)$ descendants of a quasi-primary of dimension $L$, will have two-dimensional norms in radial quantization that have the same sign as that of their primary.\footnote{Denoting by $|\OO\rangle$ the state created by the quasi-primary of dimension $L$, the norm of descendant states is given by $\langle\OO| L_1^{k+1} L_{-1}^{k+1} |\OO\rangle = (1+k)(2L+k)\langle\OO| L_1^{k} L_{-1}^{k} |\OO\rangle $ and therefore all descendant norms have the same sign as the quasi-primary norm $\langle\OO| \OO\rangle$.}

From \eqref{eq:practical_rule}, graded unitarity requires (for an operator of definite $R$-charge $R_\mathcal{O}$) 
\begin{equation}\label{eq:two-pointVirquasi}
    \mathrm{sign}\left( \kappa_{\mathcal{O}}\right) = (-1)^{L - R_{\mathcal{O}}}~.
\end{equation}
Therefore, \emph{assuming graded unitarity}, at level $L$ such a quasi-primary must contribute to the Gram matrix of inner products with sign given by
\begin{equation}\label{eq:signVirasoroquasi}
    \mathrm{sign}\left(\langle\OO| \OO\rangle\right) = (-1)^L (-1)^{L-R_{\mathcal{O}}} = (-1)^{R_{\mathcal O}}~,
\end{equation}
and $\mathfrak{sl}(2)$ descendants will also contribute \emph{with the same sign}. Consistency with graded unitarity then imposes the following requirement on the overall sign of the Kac determinant (if nonvanishing):\footnote{The Kac determinant is, of course, conventionally defined using a different basis for dimension $L$ states, but the difference will be a positive multiplicative factor.}
\begin{equation}\label{eq:Virsignprediction}
\begin{split}
    \mathrm{sign}\left(\mathrm{det}M^{(L)}\right) &= (-1)^{\sum_{R=0}^{L}R\,\dim\mathcal{V}_{R,L}}~,
\end{split}
\end{equation}
where the product runs over all states at level $L$, quasi-primary and descendant, and $\mathcal{V}_{R,L}$ refers to the graded subspace at level $L$ with given $R$ charge obtained after orthogonalization. If there are null states at level $L$ then the Kac determinant can vanish without necessarily violating graded unitarity, and in principle one would then need to investigate the corresponding reduced determinant for the simple quotient module. This will not be necessary for our purposes here.

The Kac determinant at level $L$, up to multiplication by a real, positive constant, is given by \cite{Feigin:1982tg,gorelik2007simplicity}
\begin{equation}\label{eq:KacDet}
    \mathrm{det}M^{(L)}\propto\prod_{\substack{q>p\geqslant2\\(p,q)=1}}(c-c_{p,q})^{\mathrm{dim}L^{p,q}_L}~,
\end{equation}
were $c_{p,q}$ denotes the central charge of the $(p,q)$ Virasoro minimal model,
\begin{equation}\label{eq:cpq}
    c_{p,q}=1-6\frac{(p-q)^2}{p q}~.
\end{equation}
The exponents are given by the following combinatorial formula,
\begin{equation}\label{eq:dimLpqL}
    \begin{split}
    \mathrm{dim}L^{p,q}_L&=\sum\limits_{m=1}^\infty\bigg(p_{cl}(L-(mp+1)(mq+1))+p_{cl}(L-(mp-1)(mq-1))\\
    &\qquad\qquad\quad-p_{cl}(L-(mp+1)(mq-1)-1)-p_{cl}(L-(mp-1)(mq+1)-1)\bigg)~,
    \end{split}
\end{equation}
where $p_{cl}(N)$ is the \emph{classical partition function}, which counts the number of integer partitions of $N$. Note that $\mathrm{dim}\,L^{p,q}_L=0$ for $L<(p-1)(q-1)$, from which it follows from \eqref{eq:dimLpqL} that the \emph{lowest level for which $(c-c_{p,q})$ appears as a factor in \eqref{eq:KacDet} is $L=(p-1)(q-1)$}. Note also that the factor $(c-c_{2,L+1})$ occurs with exponent \emph{exactly one} at level $L$, \emph{i.e.}, $\mathrm{dim}L^{2,L+1}_L=1$.

To illustrate the above, we have the following determinants at the first few levels:
\begin{equation}\label{eq:KacExamples}
    \begin{split}
    \mathrm{det}M^{(2)}&\propto(c-c_{2,3})=c~,\\
    \mathrm{det}M^{(4)}&\propto(c-c_{2,5})(c-c_{2,3})^2~,\\
    \mathrm{det}M^{(6)}&\propto(c-c_{2,7})(c-c_{2,5})^2(c-c_{2,3})^4(c-c_{3,4})~,\\
    \mathrm{det}M^{(8)}&\propto(c-c_{2,9})(c-c_{2,7})^2(c-c_{2,5})^4(c-c_{3,5})(c-c_{2,3})^7(c-c_{3,4})^2~.\\
    \end{split}
\end{equation}
In each expression, the linear factors are ordered so that the $c_{p,q}$'s appearing increase from left to right.

The key to imposing the unitarity requirement \eqref{eq:Virsignprediction} on the determinant formula \eqref{eq:KacDet} is the following useful combinatorial identity that we will call the \emph{Virasoro leading-term identity}:
\begin{equation}\label{eq:leadingTermIdVir}
    \sum_{\substack{q>p\geqslant2\\(p,q)=1}}\mathrm{dim}L^{p,q}_L=\sum_{R=0}^{L}R\,\dim\mathcal{V}_{R,L}~.
\end{equation}
This identity relates two expressions, each of which can be understood to give the order of the polynomial expression (in $c$) for the Kac determinant at level $L$. The fact that the left hand side gives the maximum power of $c$ is immediate from \eqref{eq:KacDet}. On the other hand, by taking an orthogonal basis for each graded subspace $\mathcal{V}_{R,L}$ and noting that the norm of any vector in $\mathcal{V}_{R,L}$ will have maximum power of $c$ given by exactly $R$ we match the right hand side of \eqref{eq:leadingTermIdVir}.

From the leading-term identity, it follows that the graded unitarity requirement \eqref{eq:Virsignprediction} amounts to the requirement that the sign of the level $L$ Kac determinant matches the parity of the number of terms in the factorized expression on the right hand side of \eqref{eq:KacDet}. In other words, \emph{only an even number of the linear factors in the product on the right hand side of \eqref{eq:KacDet} may be positive}. This will prove sufficient to rule rule out all values of $c$ not equal to $c_{2,q}$ for some $q$.

To this end, we make a number of elementary observations about the zeroes of the Kac determinant that can be observed using \eqref{eq:cpq}, the comments below \eqref{eq:dimLpqL} (and which are manifest in the examples of \eqref{eq:KacExamples}), and elementary inequalities:
\begin{itemize}
    \item $c_{2,L+1}$ is the least (\emph{i.e.}, most negative) root of \eqref{eq:KacDet} at level $L$. This implies that for $c<c_{2,L+1}$, the sign of the Kac determinant is compatible with graded unitarity for all levels up to and including $L$, as all linear factors in the Kac determinant will be negative.
    \item At level $L+1$, any new roots $c_{p,q}$ (beyond those appearing at level $L$) are \emph{greater than $c_{2,L+1}$}. This means that again, for $c<c_{2,L+1}$ all linear factors will be negative and the level $L+1$ Kac determinant will be compatible with graded unitarity.
    \item At level $L+2$, the only new root that is not larger than $c_{2,L+1}$ is $c_{2,L+3}$, which appears with multiplicity one in \eqref{eq:KacDet}. This means that if $c<c_{2,L+1}$, then compatibility with graded unitarity at level $L+2$ requires that $c\leqslant c_{2,L+3}$ as otherwise there would be a single positive factor in \eqref{eq:KacDet}.
\end{itemize}
We thus find that graded unitarity at level $L+2$ (with the given $\mathfrak{R}$-filtration) rules out any central charge value in the interval $c_{2,L+3}<c<c_{2,L+1}$, which along with the requirement $c\leqslant 0$ rules out all values of the central charge other than the $\{c_{2,q}\}$. Of course, this check is necessary but insufficient to show that the $\{c_{2,q}\}$ Virasoro VOAs are indeed graded unitary, which requires considering the norms of all states. This is expected on physical grounds as the aforementioned Argyres--Douglas SCFTs are understood to give rise to precisely these Virasoro VOAs, but would be interesting to prove rigorously.

\section{\label{sec:sl2_akm}Graded unitarity for \texorpdfstring{$\mathfrak{sl}_2$}{sl(2)} affine current algebras}

For our next study we consider putative four-dimensional theories whose associated VOA is precisely an $\mathfrak{sl}_2$ affine Kac--Moody (AKM) current algebra. These VOAs are generated by $\mathfrak{sl}_2$ currents $J^a(z)$, $a=1,2,3$, satisfying the usual OPE,
\begin{equation}\label{eq:JJOPE}
    J^a(z) J^b(0) = \frac{\frac{k}{2}\,\delta^{ab}}{z^2} + \frac{\ii f^{ab}_{\phantom{ab}c} J^c(0)}{z}+\reg~,
\end{equation}
where $f^{ab}_{\phantom{ab}c}$ are the (real) $\mathfrak{sl}_2$ structure constants and $k$ is the level, which is related to the four-dimensional flavor central charge $k_{4d}$ (the coefficient appearing in the two-point function of canonically normalized flavor currents) as \cite{Beem:2013sza}
\begin{equation}\label{eq:k2dk4d}
    k = - \frac{1}{2}k_{4d}~.
\end{equation}
The VOA of any four-dimensional $\NN=2$ SCFTs with an $\mathfrak{su}(2)$ flavor symmetry will have such a current algebra as a subalgebra, but we will be interested in possible theories $\mathcal{T}_k$ for which this is the whole story,
\begin{equation}
    \mathbb{V}[\mathcal{T}_k] \cong V_k(\mathfrak{sl}_2)~.
\end{equation}
As in the Virasoro case, there is a family of Argyres--Douglas theories (the $(A_1,D_{2n+1})$ theories) that are understood to have associated VOAs that are precisely the (simple quotient) of the $\mathfrak{sl}(2)$ current algebra at levels $k = -2 + \frac{2}{2n+1}$ \cite{Cordova:2015nma}. (These are precisely the so-called \emph{Kac--Wakimoto boundary admissible levels} for $\mathfrak{sl}_2$.)
\begin{equation}
    \mathbb{V}[{\rm AD}_{(A_1,D_{2n+1})}] \cong V_{-2 + \frac{2}{2n+1}}(\mathfrak{sl}_2)~.
\end{equation}
The elementary currents descend from Higgs branch chiral ring operators, which have vanishing $U(1)_r$ charge. As the current algebra is strongly generated by its elementary currents, it follows at once that all states are concentrated in cohomological degree zero. Since the current algebra also has integer conformal grading, the $R$-parity of all operators must be even (so the operation $s$ defined in Section \ref{sec:axiomatics} is just the identity).

The conjugation operation $\rho$ acts on the generating currents $J^a(z)$ according to
\begin{equation}
    \rho\left(J^a(z)\right)=-J^a(\zb)~.
\end{equation}
This can be obtained from \eqref{eq:conjdef} using the four-dimensional origin of the current as the flavor symmetry moment map \cite{Beem:2013sza}, but also follows from the requirement that it is an anti-linear automorphism of the VOA that commutes with the global $\mathfrak{sl}_2$ symmetry. 

\subsection{\label{subsec:sl2_R_filt}\texorpdfstring{$\mathfrak{R}$}{R}-filtration expectations}

The currents $J^a$ descend from the moment map operators in four dimensions and have $R$-charge assignment $R=1$, which coincides with their conformal weight. In the standard conformal weight-based filtration, the filtered subspace of charge $p$ is the subspace spanned by normally ordered products involving at most $p$ currents (and their derivatives). This is not the correct $\mathfrak{R}$-filtration, as the stress tensor can be written in Sugawara form,
\begin{equation}
    T=\frac{1}{k+2}\sum_{a=1}^3 \big(J^a\, J^a\big)~,
\end{equation}
but has $R[T]=1$. The simplest possibility for the $\mathfrak{R}$-filtration is to assume that this is the only ``correction'' to the conformal weight-based filtration. We therefore define a standard filtration where the $p^{\rm th}$ filtered subspace is given according to
\begin{equation}
    \mathfrak{R}_p\mathbb{V} = \mathrm{Span}\big\{L_{-m_1}\dots L_{-m_j} J^{a_1}_{-n_1}\dots J^{a_k}_{-n_k}\Omega,\ m_i\geqslant2,\,n_i\geqslant1\big\}~,\qquad j+k\leqslant p~.
\end{equation}
This defines a good filtration. (The fact that the filtration is good is readily checked as the singular terms in the $TT$, $TJ$, and $JJ$ OPEs have $R\leqslant 1$). This filtration was implicitly used in the work of \cite{Song:2016yfd}, where it was used to reproduce the Macdonald limit of the superconformal index for the aforementioned Argyres--Douglas SCFTs at low orders.

\subsection{\label{subsec:sl2_unitarity_ex}Constraints from graded unitarity at low levels}

At level $L=1$, evaluating the form \eqref{eq:hermitianform} on the $J^a$ we obtain
\begin{equation}\label{eq:kNegSl2}
    \bm{(}J^a, J^a\bm{)} = (-1)\langle J^a (\frac{\ii}{2})(-J^b(-\frac{\ii}{2})) \rangle = - k \geqslant 0~, 
\end{equation}
which reproduces the familiar requirement $k\leqslant0$ (equivalently, $k_{4d}\geqslant0$).

At level $L=2$,we consider the Sugawara vector, for which (as in the Virasoro case) the graded unitarity requirement imposes that the corresponding Virasoro central charge be negative,
\begin{equation}
    c=\frac{3k}{k+2}\leqslant0~.
\end{equation}
In the current setting, this implies that when the Sugawara vector is the true stress tensor of the VOA, then the corresponding level must be \emph{supercritical},
\begin{equation}\label{eq:kGe-2Sl2}
    k\geqslant-2~.
\end{equation}
One may proceed by constructing the quasi-primaries of higher dimension and obtain bounds similarly to what was done in Section~\ref{subsec:vir_unitarity_ex}. We will stop here and consider the more general case again by studying the sign of the Kac determinant for the vacuum module.

\subsection{\label{subsec:sl2_unitarity_kac}Only boundary admissible levels}

We denote by $k_{p,q}=-2+\frac{p}{q}$ the $(p,q)$ Kac--Wakimoto admissible level for $\mathfrak{sl}_2$, where $p\geqslant2$ and $(p,q)=1$. We will show that the sign of the determinant of the Shapovalov form on the vacuum module (due to Gorelik and Kac \cite{gorelik2007simplicity}) is not compatible with graded unitarity and the standard $\mathfrak{R}$-filtration when $k\neq k_{2,q}$.

The proof proceeds analogously to the previous Virasoro case. The essential sign constraint is that at level $L\in 2\mathbb{Z}_{\geqslant1}+1$, the sign-rule for the Gorelik--Kac determinant implies that:
\begin{equation}\label{eq:coreSl2}
    k<k_{2,L-2} \quad \Longrightarrow \quad k\leqslant k_{2,L}~.
\end{equation}
This then rules out all values of $k$ between $-2$ and $0$ other than the boundary admissible levels $\{k_{2,q}\}$, which combined with the low-level results of the previous subsection implies that only the boundary admissible levels are potentially compatible with graded unitarity (subject to the standard filtration assumption).

As in the Virasoro case, in comparing the definition of the Shapovalov form and the matrix of two-point functions in terms of which we have formulated the graded unitarity condition, there are relative signs so that the prediction of graded unitarity for the Gorelik--Kac determinant formula reads
\begin{equation}\label{eq:ChrisSignKacAKM}
    \mathrm{sign}\left(\mathrm{det} S_{L\delta+Q\alpha}(k)\right) = (-1)^{\sum_{R}R\,\dim\mathcal{V}^{Q\alpha}_{L,R}}~,
\end{equation}
where the subspaces in the sum on the right hand side are the spaces of fixed $R$-charge, conformal weight, and $\mathfrak{sl}_2$ weight in the associated graded with respect to the $\mathfrak{R}$-filtration.

\subsection*{Determinant formula}

The determinant formula for the vacuum module of the $\widehat{\mathfrak{sl}_2}$ current algebra restricted to a fixed positive affine root $\nu$ (\emph{i.e.}, a fixed conformal weight and a fixed $\mathfrak{sl}_2$ weight) is given by \cite{gorelik2007simplicity}
\begin{equation}\label{eq:KacDetAKM}
    \mathrm{det}S_\nu(k)\propto(k+2)^{m_0(\nu)}\prod_{\substack{p\geqslant2,\,q\geqslant1\\(p,q)=1}}(k-k_{p,q})^{m_{p/q}(\nu)}~,
\end{equation}
where as usual proportionality is up to an overall positive constant. The crucial combinatorial factors are the exponents $m_0(\nu)$ and $m_{p/q}(\nu)$, which are determined by the following generating functions,
\begin{equation}\label{eq:M0answer}
    \sum_{\nu\in \hat \Delta^+}m_0(\nu)e^{-\nu}=\frac{\sum_{n=1}^\infty \sum_{i=2}^\infty (e^{-i\delta})^n}{\prod_{j=1}^\infty (1-e^{-j\delta})(1-e^{-j\delta-\alpha})(1-e^{-j\delta+\alpha})}~,
\end{equation}
and
\begin{equation}\label{eq:Mp/qAnswer}
\begin{split}
    \sum_{\nu\in \hat \Delta^+} m_{p/q}(\nu)e^{-\nu}=\frac{1}{R}\bigg(\sum_{j=1}^\infty\big(&e^{-(pj-1)(qj\delta-\alpha)}+e^{-(pj+1)(qj\delta+\alpha)}\\
    &-e^{-(pj-1)qj\delta-pj\alpha}-e^{-(pj+1)qj\delta+pj\alpha}\big)\bigg)~,
    \end{split}
\end{equation}
where
\begin{equation}
    R\coloneqq(1-e^{-\alpha})\prod_{j=1}^\infty(1-e^{-j\delta})(1-e^{-j\delta-\alpha})(1-e^{-j\delta+\alpha})~.
\end{equation}
In the above, $\hat \Delta^+$ is the set of the positive affine roots, $\alpha$ is the unique positive root of $\mathfrak{sl}_2$, and $\delta$ is the imaginary root of $\widehat{\mathfrak{sl}}_2$.

From looking at the appearances of the imaginary root on the right hand side of \eqref{eq:Mp/qAnswer} one sees that whenever $(p-1)q>L$ the exponent $m_{p/q}$ will be zero at level $L$, \emph{i.e.}, for any positive affine root of the form $\nu=L\delta + n\alpha$. In particular, restricting to the ``charge-zero'' subspace ($\nu=L\delta$), we can read off the following exponents:
\begin{equation}
    \begin{split}
    m_{p/q}(L\delta)&=0~,\qquad (p-1)q > L~,\\
    m_{2/L}(L\delta)&=1~.
    \end{split}
\end{equation}
Thus, $k_{2,q}$ appears as a zero of $\det S_{L\delta}(k)$ for the first time (and with unit multiplicity) at level $L=q$. More generally, $k_{p,q}$ will appear as a zero of $\det S_{L\delta}(k)$ for the first time at level $L=(p-1)q$, though for $p\neq 2$ the multiplicity may be greater than one.

To illustrate, we display the first few (odd) level determinants in the charge-zero subspace:
\begin{equation}\label{eq:AKMKacExamples}
    \begin{split}
    \mathrm{det}S_{\delta}(k)~\propto~&(k-k_{2,1})=k~,\\
    \mathrm{det}S_{3\delta}(k)~\propto~&(k+2)^2\ (k-k_{2,3})(k-k_{2,1})^6 (k-k_{3,1})^3 (k-k_{4,1})~,\\
    \mathrm{det}S_{5\delta}(k)~\propto~&(k+2)^{12}\ (k-k_{2,5})(k-k_{2,3})^7(k-k_{3,2})^3(k-k_{2,1})^{24} \\&(k-k_{3,1})^{17} (k-k_{4,1})^8(k-k_{5,1})^3(k-k_{6,1})~.
    \end{split}
\end{equation}
At level $L=1$, compatibility of \eqref{eq:AKMKacExamples} with the unitarity requirement \eqref{eq:ChrisSignKacAKM} reproduces the familiar condition $k\leqslant0$ as in \eqref{eq:kNegSl2}. We will derive additional constraints on $k$ by imposing the unitarity requirement \eqref{eq:ChrisSignKacAKM} on the determinant formula \eqref{eq:KacDetAKM} for general odd values of $L$.

In comparing the two formulas, we will use the following key combinatorial identity:
\begin{equation}\label{eq:sumR_vs_sumMpq}
    \sum_{R=1}^L R\,\dim \mathcal{V}^{Q\alpha}_{R,L}\,=\,\sum_{\substack{p\geqslant2,\,q\geqslant1\\(p,q)=1}}m_{p/q}(L\delta+Q\alpha)~.
\end{equation}
The proof is given in Appendix~\ref{app:Lemmas}. In light of this identity, the unitarity requirement \eqref{eq:ChrisSignKacAKM} can be re-written as
\begin{equation}\label{eq:unitaryReqSl2}
    \mathrm{sign}\left(\mathrm{det}S_{L\delta+Q\alpha}\right)=(-1)^{\sum m_{p/q}(L\delta+Q\alpha)}~.
\end{equation}
Specialising to the charge-zero subspace ($Q=0$) and recalling the constraint $k\geqslant-2$ from \eqref{eq:kGe-2Sl2}, this gives us the condition
\begin{equation}\label{eq:sl2_sign_rule_product}
    \mathrm{sign}\Bigl(\prod_{\substack{p\geqslant2,\,q\geqslant1\\(p,q)=1}}(k-k_{p,q})^{m_{p/q}(L\delta)}\Bigr)=(-1)^{\sum m_{p/q}(L\delta)}~.
\end{equation}
As in the Virasoro case in Section~\ref{sec:virasoro}, we have extracted a constraint that says that only an \emph{even} number of the linear factors on the left hand side of \eqref{eq:sl2_sign_rule_product} can be positive. We can then proceed to make the following observations regarding the zeroes of the Kac--Gorelik determinant.
\begin{itemize}
    \item $k_{2,L}$ is the least (most negative) root of \eqref{eq:sl2_sign_rule_product} at level $L$. This implies that for $-2<k<k_{2,L}$ the sign of the determinant is compatible with graded unitarity for all levels up to and including $L$, as all noncritical linear factors will be negative.
    \item At level $L+1$, any new roots $k_{p,q}$ (beyond those appearing at level $L$) are \emph{larger than $k_{2,L}$}. This means that again, for $-2<k<k_{2,L}$ all noncritical linear factors will be negative and the level $L+1$ determinant will be compatible with graded unitarity.
    \item At level $L+2$, the only new root that is not larger than $k_{2,L}$ is $k_{2,L+2}$, which appears with multiplicity one. This means that if $k<k_{2,L}$, then compatibility with graded unitarity at level $L+2$ requires that $k\leqslant k_{2,L+2}$ as otherwise there would be a single positive factor in \eqref{eq:sl2_sign_rule_product}.
\end{itemize}
In conclusion, we have that compatibility of \eqref{eq:unitaryReqSl2} and \eqref{eq:KacDetAKM} requires that $k\in \{k_{2,p}\}$, \emph{i.e.}, only boundary admissible levels are potentially compatible with graded unitarity subject to our presumed $R$ filtration. Again, our checks here are insufficient to show that these VOAs are indeed graded unitary, though this is expected on the basis of their identification as the associated VOAs for the aforementioned $(A_1,D_{2p+3})$ Argyres--Douglas SCFTs.

\section{\label{sec:general_akm}Higher-rank current algebras}

In generalizing from the $\mathfrak{sl}_2$ case of the previous subsection, we will need an appropriate generalization of the standard $\mathfrak{R}$-filtration that we postulated there. On physical grounds we know that the filtration used in the $\mathfrak{sl}_2$ case will be insufficient in the higher-rank setting. This is because if a four-dimensional $\mathcal{N}=2$ SCFT has an associated vertex algebra which is an affine current algebra, then its Higgs branch must be a nilpotent orbit closure for the corresponding finite-dimensional simple Lie algebra. In this case, there will be Higgs branch relations setting to zero the generators of the ring of $\mathfrak{g}$-invariant polynomials on $\mathfrak{g}^\ast$.

To describe the consequence of these (necessary) Higgs branch relations for the $\mathfrak{R}$-filtration of the current algebra, we need to introduce a set of Casimir-type elements of $V^k(\mathfrak{g})$ that generalize the quadratic Sugawara vector. To this end, we introduce the ``Casimir operators'' of \cite{Bouwknegt:1992wg},
\begin{equation}
    C^{(s_i)}(z) = d_{a_1a_2\cdots a_{s_i}}(J^{a_1}J^{a_2}\cdots J^{a_{s_i}})(z)~,\qquad i=1,\ldots,\mathfrak{rank}(\mathfrak{g})~,
\end{equation}
where $s_i$ are the exponents of $\mathfrak{g}$ plus one, and $d_{a_1a_2\cdots a_{s_i}}$ are a basis of primitive symmetric invariant tensors for $\mathfrak{g}$.

As a consequence of the Higgs branch relations mentioned above, each such Casimir operator must belong to a component $\mathfrak{R}_pV_k(\mathfrak{g})$ of the $\mathfrak{R}$-filtration with $p<s_i$. The most modest adjustment of the naive $\mathfrak{R}$-filtration is then to assign each Casimir to the filtration component with $p= s_i-1$ (the case of the quadratic Casimir with $s=2$ just reproduces the adjustment from the previous section), so we define the following trial $\mathfrak{R}$-filtration, which is a weight-based filtration incorporating the adjustments by one to all Casimir $R$-symmetry assignments,
\begin{equation}
    \mathfrak{R}_p\mathcal{V}^{\,\bullet} = 
    \mathrm{span}\big\{
    \big(\!\!\!\prod_{i=1}^{\mathfrak{rank}(\mathfrak{g})}C^{(s_i)}_{-l^{(i)}_1}\cdots C^{(s_i)}_{-l^{(i)}_{k_i}}\big)\, J^{a_1}_{-n_1}\dots J^{a_k}_{-n_k}\Omega,\ l^{(i)}_j\geqslant s_i,~n_j\geqslant1\big\}~,\quad k+\sum_{i}(s_i-1)k_i\leqslant p\,.
\end{equation}
We conjecture that this filtration is \emph{good}, in the sense reviewed in Section~\ref{sec:axiomatics}. It should be noted that, as is well known, the OPEs of the Casimir operators do not close on the span of normally ordered products of derivatives of the Casimir operators themselves at general level (though they do for $k=1$) \cite{Bouwknegt:1992wg}. As a consequence, the goodness conjecture is not self-evidently true (at least to the authors). For the case of $\mathfrak{g}=\mathfrak{sl}_3$ it can be verified from the explicit expression for the OPEs of the cubic Casimir operator given in \cite{Bais:1987dc}, but for the higher-rank case brute force calculations will become impractical.

The conjugation automorphism $\rho$ is again determined entirely by the action $\rho(J^a)=-J^a$ on the currents (and thus for the Casimir operators we have $\rho(C^{(s_i)})=(-1)^{s_i}C^{(s_i)}$ on order-$s_i$ Casimir operators). Tracking signs between the Gram matrix and the matrix of two-point functions again leads to the following sign condition for the Kac determinant to be compatible with graded unitarity,
\begin{equation}\label{eq:ChrisSignKacAKMgeneral}
    \mathrm{sign}(\mathrm{det} S_{L\delta+\nu}) = (-1)^{\sum_{R}R\,\dim\mathcal{V}^{\nu}_{L,R}}
\end{equation}
The Gorelik--Kac determinant formula for general $\mathfrak g$ is rather involved and depends on the detailed structure of the $\mathfrak{g}$ root system. In the rest of this section we restrict to the case of simply laced $\mathfrak{g}$ (type ADE) and present specific results for $\mathfrak{g}=\mathfrak{sl}_3$ and $\mathfrak{g}=\mathfrak{sl}_4$.

The determinant formula for the $\hat{\mathfrak{g}}$ vacuum module for simply laced $\mathfrak{g}$ reads (as for $\mathfrak{sl}_2$)~\cite{gorelik2007simplicity}
\begin{equation}\label{eq:KacDetAKMgeneral}
    \mathrm{det}S_\nu(k)\propto(k+h^\vee)^{m_0(\nu)}\prod_{\substack{p\geqslant2,\,q\geqslant1\\(p,q)=1}}(k-k_{p,q})^{m_{p/q}(\nu)}~,
\end{equation}
valid up to a positive overall constant. Here, the $(p,q)$ levels are defined according to
\begin{equation}\label{eq:kpq_sec_gen}
    k_{p,q}\coloneqq-h^\vee+\frac{p}{q}~.
\end{equation}
We will refer to the above $k_{p,q}$ as \emph{minimal levels} in general, as \emph{boundary admissible} levels when $p=h^\vee$, and as \emph{inadmissible} (\emph{upper admissible}) levels when $p<h^\vee$ ($p>h^\vee$). We note that there are no inadmissible minimal levels for $\mathfrak{sl}_2$, since $p\geqslant2$, but starting from $\mathfrak{sl}_3$ inadmissible levels do appear.

The generating function of $m_0(\nu)$ reads\footnote{We are grateful to M.~Gorelik and V.~Kac for correspondence on possible simplifications of the result in~\cite{gorelik2007simplicity}. A further telescopic simplification of \eqref{eq:M0answerGeneral} is possible, but its current form is more convenient for our purposes in Appendix~\ref{app:Lemmas}.}
\begin{equation}\label{eq:M0answerGeneral}
    \sum_{\nu\in \hat \Delta^+}m_0(\nu)e^{-\nu}=\frac{1}{\tilde R}\sum_{m=2}^{h^\vee}\sum\limits_{i=m}^{\infty}  (s^{}_{m-1}-s^{}_{m})\,  \frac{e^{-i\delta}}{1-e^{-i\delta}}~,
\end{equation}
where $\delta$ is the imaginary root of $\hat{\mathfrak g}$ and $s_j\coloneqq\# \{\alpha\in \Delta^+: (\rho|\alpha)=j\}$ where $\rho$ is the Weyl vector of $\mathfrak g$ and $\Delta^+$ is the set of its positive roots, while
\begin{equation}\label{eq:Rtilde}
    \tilde{R}\coloneqq\prod_{\gamma\in\hat{\Delta}^+\setminus\Delta^+}(1-e^{-\gamma})^{\mathrm{dim}\hat{\mathfrak{g}}_{\gamma}}~,
\end{equation}
with $\hat\Delta^+$ the set of positive affine roots. Note that $\mathrm{dim}\hat{\mathfrak{g}}_{\gamma}=\mathrm{dim}{\mathfrak{h}}$ if $\gamma\in\mathbb{Z}\delta$, while $\mathrm{dim}\hat{\mathfrak{g}}_{\gamma}=1$ if $\gamma\notin\mathbb{Z}\delta$. $\tilde{R}$ is a product over only ``nonfinite'' positive affine roots.

The generating function of the exponents $m_{p/q}(\nu)$ reads
\begin{equation}\label{eq:p/q_generating_general}
    \sum_{\nu\in \hat \Delta^+} m_{p/q}(\nu)e^{-\nu}=\frac{M_{p/q}}{R}~,
\end{equation}
where
\begin{equation}\label{eq:Mp/qAnswerGeneral}
\begin{split}
    M_{p/q}&\coloneqq\sum\limits_{\alpha\in\Delta^+}\sum_{l=1}^\infty {}'\ e^{-(pl-(\rho|\alpha))ql\delta}E\bigl((pl-(\rho|\alpha))\alpha\bigr)-\sum\limits_{\alpha\in\Delta^+}\sum\limits_{l=1}^{\infty} e^{-(pl+(\rho|\alpha))ql\delta}E(pl\alpha)~,\\
    R&\coloneqq\prod_{\gamma\in\hat{\Delta}^+}(1-e^{-\gamma})^{\mathrm{dim}\hat{\mathfrak{g}}_{\gamma}}~,\\
    E(\lambda)&\coloneqq\sum\limits_{w\in {\rm Weyl}(\mathfrak{g})}(-1)^{\ell(w)} e^{w.\lambda}~,
\end{split}
\end{equation}
and where the prime on first sum in \eqref{eq:Mp/qAnswerGeneral} indicates exclusion of $l$ for which $pl-(\rho|\alpha)\leqslant0$. Here $\ell(\cdot)$ measures the length of Weyl group elements. Note that $w.\lambda=w(\lambda+\rho)-\rho$. It is shown in \cite{gorelik2007simplicity} that $M_{p/q}=0$ for $p<0$ and $p=1$. A key fact that will be used several times below is that
\begin{equation}\label{eq:Wirregular}
    E(\lambda)=0\quad \Longleftrightarrow \quad \exists \,\textbf{}\alpha\in\Delta^+~{\rm such~ that}~(\lambda+\rho|\alpha)=0~.
\end{equation}

Let us denote the coefficients appearing in the expansion of $M_{p/q}$ as follows,
\begin{equation}
    M_{p/q}=\sum\limits_{\nu\in \hat{\Delta}^+} a_{\nu} e^{-\nu}~.
\end{equation}
Now consider the restriction of this sum to a specific conformal level (coefficient of the imaginary root),
\begin{equation}
M_{p/q,L}\coloneqq\sum\limits_{\mu\in {Q}^+} a^{}_{L\delta+\mu} e^{-\mu}~,
\end{equation}
where $Q^+$ is the positive root lattice of $\mathfrak{g}$. We are particularly interested in the lowest level $L$ the factor $(k-k_{p,q})$ appears in the Gorelik--Kac determinant. Namely the minimal $L$ such that $M_{p/q,L}\not=0$. We note that at this level, the zeroes of the Gorelik--Kac determinant in all charge sectors are encoded in the functions
\begin{equation}
    \frac{M_{p/q,L}}{E(0)}~,
\end{equation}
where we have the identity $E(0)=\prod_{\alpha\in\Delta^+}(1-e^{-\alpha})$. (Note that compared with \eqref{eq:p/q_generating_general}, here the $\tilde R$ part of the denominator is omitted, because the nonfinite positive affine roots would increase the conformal level.)

From \eqref{eq:Mp/qAnswerGeneral} one can see that $M_{p/q,L}=0$ if $L$ is not divisible by $q$, and furthermore when $L=q$ we have
\begin{equation}\label{eqPq}
    M_{p/q,q}=\sum_{\alpha\in\Delta^+:(\rho|\alpha)=p-1}E(\alpha).
\end{equation}

For simply-laced $\mathfrak{g}$, we have that for $\alpha\in\Delta^+$,\footnote{We thank Maria~Gorelik for instructive correspondences on various results used below regarding ADE root systems.}
\begin{equation}\label{eq:E(theta)}
    E(\alpha)\neq0 \quad \Longleftrightarrow \quad\alpha=\theta~,
\end{equation}
where $\theta$ is the maximal root, \emph{i.e.}, $(\rho|\theta)=h^\vee-1$. Moreover, the Weyl character formula gives
\begin{equation}
    \frac{E(\theta)}{E(0)} = \ch\operatorname{Ad}\neq0~.
\end{equation}
We conclude that $M_{p/q,q}=0$ for $p\not=h^\vee$ and 
\begin{equation}
    \frac{M_{h^\vee/q,q}}{E(0)}=\ch \operatorname{Ad}~.\label{eq:chAdForBoundary_0}
\end{equation}
This means that the lowest level at which $k_{p,q}$ appears as zeroes in the Gorelik--Kac determinant is $L=q$ for $p=h^\vee$, and at least $L=2q$ for other values of $p$. In particular, $k_{p,q}$ appears as a zero at level $L=q$ for $\mathfrak{g}$-weight a root with multiplicity one for a nonzero root and multiplicity $\mathfrak{rank}(\mathfrak{g})$ at weight zero.

Recall that in the $\mathfrak{sl}_2$ case (see, \emph{e.g.}~\eqref{eq:AKMKacExamples}) we had the ``last factor'' $k-k_{2,L}$ occur with exponent one in the Cartan singlet sector $\nu=L\delta$, and could hence benefit from the sign of the determinant being correlated with the sign of $k-k_{2,L}$. Here, we see that for more general simply-laced $\mathfrak g$ we can consider nonzero roots (so affine roots $\nu=L\delta+\alpha$ with $\alpha$ a root of $\mathfrak g$) and have the factor $k-k_{h^\vee,L}$ occur with exponent one, again correlating the sign of the determinant with the sign of $k-k_{h^\vee,L}$. (Looking at the charge-zero subspace, there will only be a sign correlation when the rank of $\mathfrak{g}$ is odd.)

It is the case that $E(n\theta)\neq0$ not only for $n=1$ as used above, but for any $n\in\mathbb Z_{\geqslant0}$. It then follows from \eqref{eq:Mp/qAnswerGeneral} that for $p\geqslant h^\vee$ the lowest level $L$ where the factor $(k-k_{p,q})$ with $k_{p,q}$ a boundary or upper admissible level arises in the Gorelik--Kac determinant is $L=(p-h^\vee+1)q$. In all of these cases, the charge structure of the zeroes is determined by
\begin{equation}
    \frac{M_{p/q,(p-h^\vee+1)q}}{E(0)} = {\rm ch}\,(p-h^\vee+1){\rm Ad}~.
\end{equation}
In other words, the primitive null state appearing at level $L=(p-h^\vee+1)q$ is in the $\mathfrak{g}$ representation whose highest weight is $(p-h^\vee+1)\theta$.

For inadmissible levels ($p<h^\vee$), the general structure of the first appearance of zeros is more opaque. For a case-by-case analysis at low rank it will be useful to have the expressions for $M_{p/q,L}$ where $L$ is a low multiple of $q$ (recalling that $L$ must be a multiple of $q$),
\begin{equation}\label{eq:inadmissible_Ms}
    \begin{split}
        M_{p/q,2q}&=\!\!\!\sum_{(\rho|\alpha_+)=p-2}\!\!E(2\alpha_+)+\sum_{(\rho|\alpha_+)=2p-1}\!\!E(\alpha_+)~,\\
        M_{p/q,3q}&=\!\!\!\sum_{(\rho|\alpha_+)=p-3}\!\!E(3\alpha_+)+\sum_{(\rho|\alpha_+)=3p-1}\!\!E(\alpha_+)-\sum_{(\rho|\alpha_+)=3-p}\!\!E(p\alpha_+)~,\\
        M_{p/q,4q}&=\!\!\!\sum_{(\rho|\alpha_+)=p-4}\!\!E(4\alpha_+)+\sum_{(\rho|\alpha_+)=2p-2}\!\!E(2\alpha_+)+\sum_{(\rho|\alpha_+)=4p-1}\!\!E(\alpha_+)-\sum_{(\rho|\alpha_+)=4-p}\!\!E(p\alpha_+)~,\\
        M_{p/q,5q}&=\!\!\!\sum_{(\rho|\alpha_+)=p-5}\!\!E(5\alpha_+)+\sum_{(\rho|\alpha_+)=5p-1}\!\!E(\alpha_+)-\sum_{(\rho|\alpha_+)=5-p}\!\!E(p\alpha_+)~.
    \end{split}
\end{equation}
We recall also that $E(\lambda)$ vanishes if $\rho+\lambda$ is invariant under any Weyl reflections (\emph{i.e.}, $\rho+\lambda$ should be a regular element of $\mathfrak{h}^\ast$ for $E(\lambda)$ to be nonvanishing). These cases will turn out to be sufficient to characterize the first appearance of all inadmissible $k_{p,q}$ values for $\mathfrak{sl}_3$ and $\mathfrak{sl}_4$.

To facilitate an analysis of the constraints of graded unitarity in this setting, we will make use of a generalization of the relation \eqref{eq:sumR_vs_sumMpq}:
\begin{equation}\label{eq:sumR_vs_sumMpq_general}
    \sum_{R=1}^L R\,\dim \mathcal{V}^{\nu}_{R,L}\,=\,\sum_{\substack{p\geqslant2,\,q\geqslant1\\(p,q)=1}}m_{p/q}(L\delta+\nu)~,
\end{equation}
where $\nu$ is an element of the root lattice of $\mathfrak{g}$. The unitarity requirement \eqref{eq:ChrisSignKacAKMgeneral} then takes an analogous form to the $\mathfrak{sl}_2$ case,
\begin{equation}\label{eq:KacSignAKM}
    \mathrm{sign} \, \big(\mathrm{det}S_{L\delta+\nu}\big)=(-1)^{\sum_{\stackrel{p\geqslant2,q\geqslant1}{(p,q)=1}}m_{p/q}(L\delta+\nu)}~.
\end{equation}
Recalling also that low-level graded unitarity requires that $-h^\vee\leqslant k\leqslant 0$, we have that the critical factors $(k+h^\vee)^{m_0}$ are positive. It then follows that again, when the Gorelik--Kac determinant is nonvanishing at a given affine root, only an \emph{even} number of linear factors $(k-k_{p,q})$ in the determinant may be positive.

\subsection{\label{subsec:sl3}\texorpdfstring{$\mathfrak{sl}_3$}{sl3}}

For $\mathfrak{sl}_3$ it follows from \eqref{eq:Mp/qAnswerGeneral} and \eqref{eq:inadmissible_Ms} that the first level $L$ at which $k_{p,q}$ appears in the Gorelik--Kac determinant is given by
\begin{equation}\label{eq:sl3cases}
    L~=~\begin{cases}
        3q,\qquad &p=2~,\\
        (p-2)q~,\qquad &p\geqslant3~.
    \end{cases}
\end{equation}
For the admissible cases ($p\geqslant 3$) there is a single primitive null vector that lies in the representation with highest weight $(p-2)\theta$ (the adjoint for the boundary admissible case), while for the inadmissible cases $(p=2)$ there are two primitive null states appearing in the $[3,0]$ and $[0,3]$ representations. All have roots as weights (in the latter cases, as subsets), so all levels will appear as zeroes in the determinant evaluated at a root.

To orient ourselves, we compute the Gorelik--Kac determinant at levels $L=1,2,3,4$ and in the charge sector associated to a root $\alpha$ of $\mathfrak{sl}_3$,
\begin{equation}
\begin{adjustbox}{width=\textwidth}
$
\begin{aligned}
    \mathrm{det}S_{\delta+\alpha}(k)\,\,& \propto(k-k_{3,1})~,\\
    &\;\color{teal}{\propto (k-0)}~,\\
    \mathrm{det}S_{2\delta+\alpha}(k) & \propto (k-k_{3,2})(k-k_{3,1})^4(k-k_{4,1})^2~,\\
    &\;\color{teal}{\propto (k+\tfrac{3}{2})(k-0)^{4}(k-1)^2~,}\\
    \mathrm{det}S_{3\delta+\alpha}(k) &\propto (k+h^\vee) (k-k_{3,1})^9 (k-k_{4,1})^{18}(k-k_{3,2})^{6}(k-k_{5,1})^{3}(k-k_{2,1})^{2}~,\\
    &\;\color{teal}{\propto(k+3)(k+\tfrac{3}{2})^{6} (k+1)^{2}(k-0)^{9}(k-1)^{18}(k-2)^{3}~,}\\
    \mathrm{det}S_{4\delta+\alpha}(k) &\propto (k+h^\vee)^{5} (k-k_{3,4})(k-k_{3,2})^{19}(k-k_{2,1})^{12}(k-k_{3,1})^{15}(k-k_{4,1})^{80}(k-k_{5,1})^{22}(k-k_{6,1})^{4}~,\\
    &\;\color{teal}{\propto (k+3)^{5} (k+\tfrac{9}{4})(k+\tfrac{3}{2})^{19}(k+1)^{12}(k-0)^{15}(k-1)^{80}(k-2)^{22}(k-3)^{4}~.}
\end{aligned}
$
\end{adjustbox}
\end{equation}
From these one concludes that, given the sign rule arising from graded unitarity, at level $L=1$ one requires $k\leqslant0$, and level $L=2$ one requires that if $k<0$ then $k\leqslant k_{3,2}$, at level $L=3$ one finds no new constraint, and at level $L=4$ one requires if $L<k_{3,2}$ then $k\leqslant k_{3,4}$.

The pattern is reminiscent of what we saw for the $\mathfrak{sl}_2$ case, and as in that case, the argument can be generalized to all levels. The key facts about the zeroes in the $\mathfrak{sl}_3$ Gorelik--Kac determinant (evaluated at a root)---which can be verified from the above expressions for levels where roots first appear via elementary manipulation of inequalities---are as follows:
\begin{itemize}
    \item At level $L=3n+1$, the two least (most negative) noncritical roots are $k_{3,3n+1}$ (with unit multiplicity) and $k_{3,3n=1}$.
    \item At level $L=3n+2$, the two least noncritical roots are $k_{3,3n+2}$ (with unit multiplicity) and $k_{3,3n+1}$. 
    \item At level $L={3n+3}$, the least noncritical root is again $k_{3,3n+2}$.
\end{itemize}
Combined with the sign rule, this analysis rules out all levels other than the boundary admissible levels $k_{3,q}$, subject to the presumed $\mathfrak{R}$-filtration.

\subsection{\label{subsec:sl4}\texorpdfstring{$\mathfrak{sl}_4$}{sl4}}

For $\mathfrak{sl}_4$ we can again determine the first levels at which the various roots of the Gorelik--Kac determinant arise. From \eqref{eq:Mp/qAnswerGeneral} and \eqref{eq:inadmissible_Ms} we have the following first levels,
\begin{equation}\label{eq:sl4cases}
    L=\begin{cases}
        2q~,\qquad &p=2,3~,\\
        (p-3)q~,\qquad &p\geqslant4~.
    \end{cases}
\end{equation}
For the inadmissible levels $k_{2,q}$, the corresponding primitive null vector is always in the adjoint representation, while for the inadmissible $k_{3,q}$ levels the primitive null vector is in $[0,2,0]$ representation. For the admissible levels, the primitive null has highest weight $(p-3)\theta$ (\emph{i.e.}, it is the $[(p-3),0,(p-3)]$ representation). All of these representations have the roots of $\mathfrak{sl}_4$ as weights, so these levels will all appear as zeroes of the determinant evaluated as a root.

We can largely proceed as in the lower-rank cases. Here, the key facts about the order of appearance of zeroes in the Gorelik--Kac formula (in the charge-sector corresponding to a root) are as follows:
\begin{itemize}
    \item At level $L=4n+1$ with $n\geqslant 1$, the two least (most negative) noncritical roots are $k_{4,4n+1}$ (with unit multiplicity) and $k_{4,4n-1}$.
    \item At level $L=4n+2$ with $n\geqslant 1$, the two least noncritical roots are $k_{2,2n+1}$ (with unit multiplicity) and $k_{4,4n+1}$. 
    \item At level $L=4n+3$, the two least noncritical roots are $k_{4,4n+3}$ (with unit multiplicity) and $k_{2,2n+1}$.
    \item At level $L=4n$ with $n\geqslant 1$, the least noncritical root is again $k_{4,4n+3}$.
\end{itemize}
Along with the Sugawara requirement that $k\geqslant -4$, this information allows us to exclude all values of $k$ other than the boundary admissible levels $k_{4,q}$ and the inadmissible levels $k_{2,2n+1}$ in the range $k\leqslant k_{2,1}=-2$.

Dealing with very small values of $k$ requires a bit of bespoke analysis. At levels one and two (still in a root sector) we have
\begin{equation}
\begin{split}
    \mathrm{det}S_{\delta+\alpha}(k)&\propto (k-k_{4,1})~,\\
    &\;\color{teal}{\propto (k-0)~,}\\
    \mathrm{det}S_{2\delta+\alpha}(k)&\propto (k-k_{2,1})(k-k_{3,1})(k-k_{4,1})^{m_{4/1}(2\delta+\alpha)}(k-k_{5,1})^{m_{5/1}(2\delta+\alpha)}~,\\
    &\;\color{teal}{\propto (k+2)(k+1)(k-0)^{m_{4/1}(2\delta+\alpha)}(k-1)^{m_{5/1}(2\delta+\alpha)}~,}\\
\end{split}
\end{equation}
By our familiar sign rule, levels $-2<k<-1$ are excluded from the level-one determinant. At level two, there is a null state giving rise to the zero at $k_{3,1}=-1$, but this is a component of the $[0,2,0]$ representation, whereas the zero at $k=k_{2,1}=-2$ is a component of the adjoint representation. Thus, by passing to a representation basis rather than a weight basis for our states we further exclude the full range $-2<k<-1$ including $k=k_{3,1}$. 

The final result of this analysis is that besides the boundary admissible levels $k_{4,q}$, also the inadmissible levels $k_{2,q}$ are allowed by the GK determinant considerations to the extent that we have probed them. In actuality, while the boundary admissible levels are understood to arise from generalised Argyres--Douglas theories \cite{Xie:2016evu}, the $(2,q)$ inadmissible levels likely do not. For the first case of $k_{2,1}=-2$, in fact, it is known that the associated variety is a Dixmier sheet rather than a nilpotent orbit closure \cite{Arakawa_Moreau:2017157,arakawa2024generalized}, and at this value of the level the $\mathfrak{sl}_4$ current algebra is not quasi-lisse. It would be of considerable interest to discern whether a more granular application of the requirements of graded unitarity at these levels could rule them out (subject to the usual assumption for the $\mathfrak{R}$-filtration).


\acknowledgments

The authors thank Gonenc Mogol and Balt van Rees for collaboration in the initial stages of this project. The authors would also like to especially thank M.~Gorelik for correspondence regarding the determinant formula for vacuum modules of affine current algebras. The authors are grateful to T.~Arakawa, D.~Butson, A.~Deb, N.~Garner, V.~Kac, J.~Song, M.~Tuite, and W.~Yan for useful discussions and collaborations on related topics. The work of CB was supported in part by grant \#494786 from the Simons Foundation, by ERC Consolidator Grant \#864828 ``Algebraic Foundations of Supersymmetric Quantum Field Theory'' (SCFTAlg), and by the STFC consolidated grant ST/X000761/1. The work of AA and LR was supported in part by the National Science Foundation under Grant NSF PHY-2210533 and by the Simons Foundation under Grant 681267 (Simons Investigator Award). ML is supported in part by the STFC Consolidated grant ST/X000591/1.


\vfill\eject

\appendix


\section{\label{app:conventions}Superconformal algebra and multiplets}

In this appendix we spell out our conventions and summarize the four-dimensional $\mathcal{N}=2$ superconformal algebra. We follow the conventions of \cite{Bianchi:2019sxz} with a swap in the signs of $\sigma^4$ and $\bar{\sigma}^4$ (which leads to a different definition for $z$ in terms of $x^3$ and $x^4$), and with $\qq_{\,2}^{(\zeta)}= -\zeta \qq_{\,2}^{\mathrm{\;\;there}} $. We raise and lower $\mathfrak{su}(2)$ indices according to $\phi^a = \epsilon^{a b} \phi_b$, $\phi_a = \epsilon_{ab} \phi^b$, where $\epsilon_{12}=1$ and $\epsilon^{12}=-1$. Our sigma matrices are taken to be
\begin{equation}
    \sigma^{\mu}_{\aa \bbd} =(\sigma^a,-\ii \mathbbm{1})~,\qquad  (\bar \sigma^{\mu})^{ \aad \bb} =(\sigma^a,\ii \mathbbm{1})~,
\end{equation}
where $\sigma^a$ are the Pauli matrices, $\alpha= \pm$, and $\dot{\alpha}= \dot{\pm}$.

\subsection{\texorpdfstring{$R$}{R}-symmetry algebra}

The generators $\RR^\II_{\phantom{1}\JJ}$ of the $\mathfrak{u}(2)$ $R$-symmetry algebra obey the commutation relations
\begin{equation}
    [\RR^\II_{\phantom{\II}\JJ},\RR^{\mathcal{K}}_{\phantom{\KK}\LL}]=\delta^\mathcal{K}_{\phantom{\KK}\JJ}\RR^\II_{\phantom{\II}\LL}-\delta^\II_{\phantom{\II}\LL}\RR^\mathcal{K}_{\phantom{\KK}\JJ}~.
\end{equation}
We define the $\mathfrak{su}(2)_R$ generators ($\RR^+$, $\RR^-$ and $\RR$) and the $\mathfrak{u}(1)_r$ generator ($r$) from $\RR^\II_{\phantom{\II}\JJ}$ according to
\begin{equation}
    \RR^1_{\phantom{1}2}=\RR^+~,\qquad\RR^2_{\phantom{2}1}=\RR^-~,\qquad\RR^1_{\phantom{1}1}=\frac12 r+\RR~,\qquad\RR^2_{\phantom{1}2}=\frac12 r-\RR~,
\end{equation}
so the $\mathfrak{su}(2)_R$ generators obey the standard algebra
\begin{equation}
    [\RR^+,\RR^-]=2\RR~,\qquad [\RR,\RR^{\pm}]=\pm\RR^{\pm}~.
\end{equation}
We sometimes use the generators $R_1= \frac{1}{2}(R_{+}+R_{-})$, $R_2= \frac{\ii}{2}(R_{-}-R_{+})$. We denote the eigenvalue of $\RR$ by $R$, and the eigenvalue of $r$ by the same letter.

\subsection{\label{subapp:superalgebra}Superconformal algebra}

The generators of the conformal algebra are $P^\mu$, $K^\mu$, $D$ and $M_{\mu \nu}$ and we denote the eigenvalue of operators under $D$ by $E$.\footnote{Here, following \cite{Simmons-Duffin:2016gjk}, we adopt conventions where the generators of the conformal algebra are \emph{anti-}Hermitian.} We write generators with spinor indices instead of vector indices, so $P_{\aa \aad} \coloneqq \sigma^\mu_{\aa \aad} P_\mu$, $K^{ \aad \aa} \coloneqq \bar{\sigma}_\mu^{\aad \aa} K^\mu$, and
\begin{equation}
    \MM_{\bb}^{\phantom{\bb} \aa}\coloneqq -\tfrac14 \bar{\sigma}^{\mu \aad  \aa} \sigma^\nu_{ \bb \aad} M_{\mu \nu}~,\qquad
    \MM^{\aad}_{\phantom{\aad} \bbd}\coloneqq - \tfrac14  \bar{\sigma}^{\mu \aad \aa} \sigma^\nu_{ \aa \bbd} M_{\mu \nu}~.
\end{equation}
The half-integer-valued spins $j_1$ and $j_2$ are the eigenvalues of $\MM_{+}^{\phantom{\bb} +}$ and $\MM_{\pd}^{\phantom{\bb} \pd}$ respectively.

In these terms, the $4d$ $\NN=2$ superconformal algebra is given as follows:
\begingroup
\allowdisplaybreaks
\begin{alignat*}{6}
    &[\MM_{\aa}^{~\bb},\MM_{\gamma}^{\phantom{\gamma}\delta}] &~=~&~\delta_{\gamma}^{~\bb}\MM_{\aa}^{~\delta}-\delta_{\aa}^{~\delta}\MM_{\gamma}^{~\bb}~, \qquad
    &&[\MM^{\aad}_{~\bbd},\MM^{\ggd}_{~\ddd}] &~=~&~\delta^{\aad}_{~\ddd}\MM^{\ggd}_{~\bbd}-\delta^{\ggd}_{~\bbd}\MM^{\aad}_{~\ddd}~,\\
    &[\MM_{\aa}^{~\bb},\PP_{\gamma\ggd}]&~=~&~\delta_{\gamma}^{~\bb}\PP_{\aa\ggd}-\tfrac12\delta_{\aa}^{\phantom{\aa}\bb}\PP_{\gamma\ggd}~, \qquad
    &&[\MM^{\aad}_{~\bbd},\PP_{\gamma\ggd}] &~=~&~\delta^{\aad}_{~\ggd}\PP_{\gamma\bbd}-\tfrac12\delta^{\aad}_{\phantom{\aad}\bbd}\PP_{\gamma\ggd}~,\\
    &[\MM_{\aa}^{~\bb},\KK^{\ggd\gamma}] &~=~&~	-\delta_{\aa}^{~\gamma}\KK^{\ggd\bb}+\tfrac12\delta_{\aa}^{\phantom{\aa}\bb}\KK^{\ggd\gamma}~,\qquad
    &&[\MM^{\aad}_{~\bbd},\KK^{\ggd\gamma}] &~=~&~	-\delta^{\ggd}_{~\bbd}\KK^{\aad\gamma}+\tfrac12\delta^{\aad}_{\phantom{\aad}\bbd}\KK^{\ggd\gamma}~,\\
    &[\HH,\PP_{\aa\aad}] &~=~&~	\PP_{\aa\aad}~,\qquad
    &&[\HH,\KK^{\aad\aa}] &~=~&~	- \KK^{\aad\aa}~,\\
    &[\KK^{\aad\aa},\PP_{\bb\bbd}] &~=~&~	4 \delta_{\bb}^{\phantom{\bb}\aa}\delta^{\aad}_{\phantom{\aad}\bbd}\HH+4\delta_{\bb}^{\phantom{\bb}\aa}\MM^{\aad}_{\phantom{\aad}\bbd}+4\delta^{\aad}_{\phantom{\aad}\bbd}\MM_{\bb}^{\phantom{\bb}\aa}~,&&~ & &\\
    &[\MM_{\aa}^{~\bb},Q_{\gamma}^\II]	&~=~&~	\delta_{\gamma}^{~\bb} Q_{\aa}^\II -\tfrac12\delta_{\aa}^{\phantom{\aa}\bb} Q_{\gamma}^\II~,	\qquad
    &&[\MM^{\aad}_{~\bbd},\tilde{Q}_{\II \ddd}] &~=~&~ \delta^{\aad}_{~\ddd}\tilde{Q}_{\II \bbd} -\tfrac12\delta^{\aad}_{\phantom{\aad}\bbd}\tilde{Q}_{\II \ddd}~,\\
    &[\MM_{\aa}^{~\bb},S_{\II}^{\phantom{\aa}\gamma}] &~=~&~	-\delta_{\aa}^{~\gamma}S_{\II}^{\phantom{\aa}\bb}+\tfrac12\delta_{\aa}^{\phantom{\aa}\bb} S_{\II}^{\phantom{\aa}\gamma}~,
    \qquad &[&\MM^{\aad}_{~\bbd},\tilde{S}^{\II\ggd}] &~=~&~	-\delta^{\ggd}_{~\bbd}\tilde{S}^{\II\aad}+\tfrac12\delta^{\aad}_{\phantom{\aad}\bbd}\tilde{S}^{\II\ggd}~,\\
    &[\HH,Q_{\aa}^\II] &~=~&~	\tfrac12 Q_{\aa}^\II~,\qquad
    &&[\HH,\tilde{Q}_{\II \aad}] &~=~&~ 	\tfrac12 \tilde{Q}_{\II \aad}~,\\
    &[\HH, S_{\II}^{\phantom{\aa}\aa}] &~=~&~	-\tfrac12  S_{\II}^{\phantom{\aa}\aa}~,\qquad
    &&[\HH, \tilde{S}^{\II\aad} ] &~=~&~	-\tfrac12 \tilde{S}^{\II\aad} ~,\\
    &[\RR^\II_{\phantom{\II}\JJ},Q_{\aa}^\mathcal{K}]&~=~&	\delta_{\JJ}^{~\mathcal{K}} Q_{\aa}^\II -\frac{1}{4} \delta_{\JJ}^{\II} Q_{\aa}^\mathcal{K}~,\qquad
    &&[\RR^\II_{\phantom{\II}\JJ},\tilde{Q}_{\mathcal{K} \aad}]	&~=~&~	-\delta_{\mathcal{K}}^{~\II} \tilde{Q}_{\JJ \aad} +\frac{1}{4} \delta_{\JJ}^{\II} \tilde{Q}_{ \mathcal{K} \aad}~,\\
    &[\RR^\II_{\phantom{\II}\JJ},\tilde{S}^{\mathcal{K} \aad}]&~=~&	\delta_{\JJ}^{~\mathcal{K}} \tilde{S}^{\II\aad} -\frac{1}{4} \delta_{\JJ}^{\II} \tilde{S}^{\mathcal{K}\aad}~,\qquad
    &&[\RR^\II_{\phantom{\II}\JJ},S_{\mathcal{K}}^{\aad}]	&~=~&~	-\delta_{\mathcal{K}}^{~\II} S_{\JJ}^{ \aad} +\frac{1}{4} \delta_{\JJ}^{\II} S_{ \mathcal{K}}^{\aad}~,\\
    &[\KK^{\aad\aa},Q_{\bb}^\II] &~=~&~	2 \delta_{\bb}^{\phantom{\bb}\aa}\tilde{S}^{\II\aad}~,\qquad
    &&[\KK^{\aad\aa},\tilde{Q}_{\II \bbd}] &~=~&~	2 \delta_{\bbd}^{\phantom{\bbd}\aad} S_{\II}^{\phantom{\aa}\aa}~,\\
    &[\PP_{\aa\aad},S_{\II}^{\phantom{\aa}\bb}]	&~=~&~	- 2 \delta_{\aa}^{\phantom{\aa}\bb}\tilde{Q}_{\II \aad}~,\qquad
    &&[\PP_{\aa\aad},\tilde{S}^{\II\bbd} ] &~=~&~	-2 \delta_{\aad}^{\phantom{\aad}\bbd} Q_{\aa}^\II~,\\
    &\{Q_{\aa}^\II,\,\tilde{Q}_{\JJ\aad}\} &~=~&\tfrac12	\delta^\II_{\phantom{\II}\JJ} \PP_{\aa\aad}~,\qquad
    &&\{\tilde{S}^{\II\aad},\,S_{\JJ}^{\phantom{\aa}\aa}\} &~=~&	\tfrac12 \delta^\II_{\phantom{\II}\JJ} \KK^{\aad\aa}~,\\
    &\{Q_{\aa}^\II,\,S^{\phantom{\aa}\bb}_\JJ\} &~=~&	\tfrac12 \delta^\II_{\phantom{\II}\JJ}\delta_{\aa}^{\phantom{\aa}\bb}\HH   + \delta^\II_{\phantom{\II}\JJ} \MM_{\aa}^{\phantom{\aa}\bb}-\delta_\aa^{\phantom{\aa}\bb} \RR^\II_{\phantom{\II}\JJ}~,&&~ & &\\
    &\{\tilde{S}^{\II\aad},\,\tilde{Q}_{\JJ\bbd}\}		&~=~&	\tfrac12 \delta^\II_{\phantom{\II}\JJ}\delta^{\aad}_{\phantom{\aad}\bbd}\HH + \delta^\II_{\phantom{\II}\JJ} \MM^{\aad}_{\phantom{\aad}\bbd}+\delta^{\aad}_{\phantom{\aad}\bbd} \RR^\II_{\phantom{\II}\JJ}~.&&~ & &
\end{alignat*}
\endgroup

\subsection{\label{subapp:VOA_conventions}Chiral subalgebra conventions}

The generators of the $\mathfrak{sl}(2)\times \overline{\mathfrak{sl}(2)}$ conformal subalgebra that acts in the chiral algebra plane are defined as
\begin{equation}\label{eq:2dsl2}
\begin{alignedat}{4}
    &2 L_{-1}\coloneqq \PP_{+ \pd} = P_{3} - \ii P_4~, \qquad  2 L_{+1}& \coloneqq  \KK^{\pd +}=K_{3} + \ii K_4 ~, \qquad 2 L_0&\coloneqq  \HH + \MM~, \\
    &2 \Lb_{-1}\coloneqq -\PP_{- \md} =  P_{3} + \ii P_4 ~, \qquad 2 \Lb_{+1} &\coloneqq  - \KK^{\md -}= K_{3} - \ii K_4 ~, \qquad 2 \Lb_0&\coloneqq  \HH - \MM~,
\end{alignedat}
\end{equation}
where $\MM$ are rotations in the chiral algebra plane
\begin{equation}
    \MM \coloneqq \MM_{+}^{\phantom{+}+} + \MM_{\pd}^{\phantom{\pd}\pd}~.
\end{equation}
The complex coordinates on the chiral algebra plane are then given by
\begin{equation}\label{eq:zzbtox}
    z\coloneqq x^3 + \ii x^4~, \qquad \bar{z}\coloneqq x^3 - i x^4~.
\end{equation}
In planar quantization with Euclidean time coordinate $x_4$, we have that $L_{-1}^{\dagplane} = - L_{-1}$ and that $\Lb_{-1}^{\dagplane} = - \Lb_{-1}$.

The twisted $\widehat{\mathfrak{sl}}(2)$ generated by 
\begin{equation}
    \Lh_{-1}\coloneqq \Lb_{-1} - \zeta R_-~, \qquad \Lh_{+1}\coloneqq \Lb_{+1} +\frac{1}{\zeta} R_+~, \qquad \Lh_0 \coloneqq \Lb_0 - \RR~,
\end{equation}
is exact with respect to $\qq_{\,1}$ and $\qq_{\,2}$ given by
\begin{equation}
    \qq_{\,1}^{(\zeta)} = \mathcal{Q}^1_{-}+\zeta \widetilde{\mathcal{S}}^{2\dot{-}}~,\qquad \qq_{\,2}^{(\zeta)} = \widetilde{\mathcal{Q}}_{2\dot{-}}-\zeta \mathcal{S}_1^{-}~.
\end{equation}

\subsection{\label{app:multiplets}Superconformal multiplets}

We follow the conventions of \cite{Dolan:2002zh} for the naming of representations of the four-dimensional $\mathcal{N}=2$ superconformal algebra. Of relevance to this work are the superconformal multiplets containing Schur operators, which are listed in Table~\ref{tab:schurmult}. Each of these multiplets contains exactly one Schur operator, which will be a quasi-primary in the VOA.

\renewcommand{\arraystretch}{1.5}
\begin{table}[t]
\centering
\begin{tabular}{|l|l|l|l|l|}
\hline \hline
Multiplet  & $\OO_{\rm Schur}(0)$  & $h$ & $r$  & $R_{\OO_{\rm Schur}}$ \\ 
\hline 
$\hat \BB_R$  &  $\Psi^{11\dots 1}(0)$   &    $R$ &  $0$ & $R$ \\ 
\hline
$\DD_{R (0, j_2 )}$  &    $ \widetilde {\QQ}^1_{\dot +} \Psi^{11\dots 1}_{\dot  + \dots \dot  + }(0)$ &   $R+ j_2 +1$  & $j_2 + \frac{1}{2}$  & $R+\frac{1}{2}$ \\
\hline
$\bar \DD_{R (j_1, 0 )}$  & $ {\QQ}^1_{ +} \Psi^{11\dots 1}_{+   \dots +}(0)$ &     $R+ j_1 +1$  & $-j_1 - \frac{1}{2}$  & $R+\frac{1}{2}$ \\
\hline
$\hat \CC_{R (j_1, j_2) }$ &   ${\QQ}^1_{+} \widetilde  {\QQ}^1_{\dot +} \Psi^{11\dots 1}_{+   \dots + \, \dot  + \dots \dot  + }(0)$&   
$R+ j_1 + j_2 +2$  & $j_2 - j_1$  &
$R+1$ \\
\hline
\end{tabular}
\caption{\label{tab:schurmult}Superconformal multiplets containing Schur operators. $\Psi$ denotes the superconformal primary of the relevant multiplet, the second column illustrates how the Schur operator is obtained from the superconformal primary, as well as the quantum numbers $h$, $r$ and $R_{\OO_{\rm Schur}}$ of the Schur operator in terms of the quantum numbers of $R$, $j_1$, $j_2$ of the superconformal primary.}
\end{table}

Finally we collect a small selection of OPE selection rules from \cite{Nirschl:2004pa,Liendo:2015ofa,Lemos:2015orc} that are relevant for tracking $R$ charges in the chiral algebra OPE. We show only multiplets containing Schur operators on the right hand side:
\begin{align}\label{eq:OPEselrules}
\begin{split}
    \hat{\BB}_{R_1} \times \hat{\BB}_{R_2} &~\sim~ \sum\limits_{R=R_2-R_1}^{R_1+R_2} \hat{\BB}_R+\sum_{j \in \frac{\mathbb{N}_0}{2}} \sum\limits_{R=R_2-R_1}^{R_1+R_2-1} \hat{\CC}_{R(j,j)}~,\\
    \hat{\CC}_{0(0,0)} \times \hat{\CC}_{0(0,0)} &~\sim~ \mathbf{1}+\sum_{j \in \frac{\mathbb{N}_0}{2}}\hat{\CC}_{0(j,j)}+\sum_{j \in \frac{\mathbb{N}_0}{2}}\hat{\CC}_{1(j,j)}~,\\
    \hat{\CC}_{0(0,0)} \times \hat{\BB}_1 &~\sim~ \hat{\BB}_1+\hat{\BB}_2+\sum_{j \in \frac{\mathbb{N}_0}{2}}\hat{\CC}_{0(j,j)}+\sum_{j \in \frac{\mathbb{N}_0}{2}} \hat{\CC}_{1(j,j)}~.
\end{split}
\end{align}
Here $\hat{\BB}_0=\mathbf{1}$ is identified with the identity operator.

\section{\label{app:Dagger}Two point functions and radial quantization norms}

To determine the requirements for Kac determinants based on the sign requirements of graded unitarity as presented in the main text of this paper, one must compare the (signs of) norms of states defined in radial quantization (the Gram matrix/Shapovalov form, for which determinant formulae are available) with the coefficients appearing in the two-point functions of the corresponding operators. In the language of two-dimensional conformal field theory, the radial adjoint of a two-dimensional quasi-primary vertex operator $\mathcal{O}\in\mathcal{V}$ with weight $h$ is given by
\begin{equation}\label{eq:dagradO}
    \left(\OO(z)\right)^\dagrad = (-1)^h \zb^{-2h} \phi(\OO)\left(\frac{1}{\zb}\right)~,
\end{equation}
where $z^*=\zb$ and $\phi:\mathcal{V}\to\mathcal{V}$ is an anti-unitary involution (the avatar of taking the adjoint of the operator in Lorentzian signature; we will later adopt the notation $\phi(\mathcal{O})=\mathcal{O}^\dagger$). Writing the mode expansion of $\OO(z)$ as\footnote{Note that here we are using the physics mode convention which differs from the math convention used in section~\ref{sec:intro} as $\OO_{-h}= a_{\OO_{-1}}$.} 
\begin{equation}
    \OO(z) = \sum\limits_{n} \OO_{n}z^{-h-n}~,
    \label{eq:modeexp}
\end{equation}
this gives the following conjugation action on modes of a quasi-primary,
\begin{equation}
    \left(\OO_n\right)^\dagrad = (-1)^h (\phi(\OO))_{-n}~.
\end{equation}
For the stress tensor $T(z)$ with $h=2$ one has $\phi(T)=T$, and correspondingly the mode convention
\begin{equation}
    (L_n)^\dagger = L_{-n}~.
\end{equation}
For affine currents $J(z)$ with $h=1$ in the standard physics normalization, one has $\phi(J^a)=-J^a$, leading again to the natural conjugation action on modes,
\begin{equation}
    (J_n)^\dagger = J_{-n}~.
\end{equation} 
For quasi-primary operators, one then has the relation
\begin{equation}\label{eq:norm_vs_2pt_quasiprimary}
    \langle \mathcal{O} | \mathcal{O} \rangle = (-1)^h \langle \Omega | (\phi(\mathcal{O})_{+h}\mathcal{O}_{-h}|\Omega\rangle = (-1)^h\lim_{z\to\infty} z^{2h}\langle \phi(\mathcal{O})(z)\mathcal{O}(0)\rangle~,
\end{equation}
so the Gram norm of the state $|\mathcal{O}\rangle$ is read off from the constant term in the $\phi(\mathcal{O})\mathcal{O}$ two-point function up to a factor of $(-1)^h$.

The relation between two-point functions and radial quantization norms is slightly more elaborate for descendants. Let us denote the $m^{\rm th}$ derivative of $\mathcal{O}$ by $\mathcal{O}^{(m)}$. Then the conjugate of $\mathcal{O}^{(m)}$ will consist of a linear combination of up to $m$ derivatives of $\mathcal{O}^\dagger = \phi(\mathcal{O})$,
\begin{equation}\label{eq:paramDagger}
\begin{split}
  \big(\mathcal{O}^{(m)}(z)\big)^\dagger=(-1)^{h+m}\left(a_0^{(m)}\mathcal{O}^{\dagger(m)}(\tfrac{1}{\bar{z}})\,\bar{z}^{-2h-2m}+a_1^{(m)}\mathcal{O}^{\dagger(m-1)}\right.&(\tfrac{1}{\bar{z}})\,\bar{z}^{-2h-2m+1}+\cdots\\
  &\left.\cdots+a_m^{(m)}\mathcal{O}^{\dagger}(\tfrac{1}{\bar{z}})\,\bar{z}^{-2h-m}\right)~,
\end{split}
\end{equation}
where $h$ is the conformal weight of the quasi-primary $\mathcal{O}$. Using the relation
\begin{equation}
    \big(O^{(m+1)}(z)\big)^\dagger = \partial_{\bar{z}} \big( O^{(m)}(z)\big)^\dagger~,
\end{equation}
(which follows, \emph{e.g.}, from replacing the operators with their mode expansions) implies the recurrence relation
\begin{equation}
    a^{(m+1)}_j=a^{(m)}_j+(2h+2m-j+1)a^{(m)}_{j-1}~,
\end{equation}
with $a_0^{(m)}=1$, as well as $a_j^{(m)}=0$ for $j>m$. This can be solved to find
\begin{equation}
    a^{(m)}_j=\frac{m!}{(m-j)!j!}\cdot\prod_{i=1}^j(2h+m-i)~,
\end{equation}
which can be resummed nicely to give the general relation (for arbitrary states),
\begin{equation}
    \mathcal{O}(z)^\dagger = \left(e^{\bar{z}L_{+1}}(-\bar{z}^{-2})^{L_0}\phi(\mathcal{O})\right)(\tfrac{1}{\bar{z}})~.
\end{equation}

Now, if we consider the two-point function
\begin{equation}
    \langle \phi(\mathcal{O})(z)\mathcal{O}(w)\rangle=\frac{\kappa_\mathcal{O}
    }{(z-w)^{2h}}~,
\end{equation}
then taking derivatives gives
\begin{equation}\label{eq:derO2pt}
    \langle \phi(\mathcal{O})^{(m-j)}(z)\mathcal{O}^{(m)}(w)\rangle=(-1)^{m-j}\,\frac{2h(2h+1)\cdots(2h+2m-j-1)\, \kappa_\mathcal{O}}{(z-w)^{2h+2m-j}}~.    
\end{equation}
Thus we have for descendants,
\begin{equation}\label{eq:GramNormSimp}
\begin{split}
    \langle\big(\mathcal{O}^{(m)}(z)\big)^\dagger\,\mathcal{O}^{(m)}(0)\rangle&=\sum_{j=0}^m (-1)^{h+m}\, a^{(m)}_j\, \bar{z}^{-2h-2m+j}\,\langle \mathcal{O}^{(m-j)}\big(\tfrac{1}{\bar{z}}\big)\mathcal{O}^{(m)}(0)\rangle\\
    &=\sum_{j=0}^m \frac{(-1)^{h+j} m!}{(m-j)!j!}\frac{(2h+m-1)!}{(2h+m-j-1)!}\frac{(2h+2m-j-1)!}{(2h-1)!}\kappa_\mathcal{O}~,\\
    &=(-1)^h\frac{m!(2h+m-1)!}{(2h-1)!}\kappa_{\mathcal{O}}~.
\end{split}
\end{equation}
Comparing \eqref{eq:GramNormSimp} and \eqref{eq:derO2pt} (with $j=0$) we have the following relation between Gram norms and two-point functions (generalizing \eqref{eq:norm_vs_2pt_quasiprimary}),
\begin{equation}\label{eq:norm_vs_2pt_descendant}
    \langle\mathcal{O}^{(m)}|\mathcal{O}^{(m)}\rangle = (-1)^{h+m}\frac{(2h+m-1)!}{(2h+2m-1)!}\lim_{z\to\infty}z^{2h+2m}\langle\phi(\mathcal{O})^{(m)}(z)\mathcal{O}^{(m)}(0)\rangle~.
\end{equation}

\section{\label{app:Lemmas}Proofs of combinatorial identities}

In this appendix, we first prove the combinatorial identity \eqref{eq:sumR_vs_sumMpq} for $\mathfrak{sl}_2$ affine current algebras. In the second part, we present the argument for the more general version \eqref{eq:sumR_vs_sumMpq_general} that applies for $\mathfrak{sl}_N$ current algebras, and briefly address a generalization to arbitrary simple $\mathfrak{g}$, though those cases are not discussed in the main text.

\subsection{Exponent identity for \texorpdfstring{$\mathfrak{sl}_2$}{sl(2)}}

To derive \eqref{eq:sumR_vs_sumMpq} we will make use of the conformal weight-based filtration $\mathfrak{W}_\bullet$ in addition to the (postulated) $\mathfrak{R}$-filtration defined on the generic-level (or alternatively, universal) current algebra $V^k(\mathfrak{sl}_2)$. Define the associated graded vertex Poisson algebras with respect to both filtrations and their corresponding decomposition into homogeneous subspaces as
\begin{equation}
\begin{split}
    {\rm gr}^{\mathfrak W}V^k(\mathfrak{sl}_2) &= \bigoplus_{h,w}V_{h,w}~,\\
    {\rm gr}^{\mathfrak R}V^k(\mathfrak{sl}_2) &= \bigoplus_{h,R}\mathcal{V}_{h,R}~.
\end{split}
\end{equation}
When we wish to specify a given charge sector ($\mathfrak{sl}_2$ weight) $Q$, we write $V_{h,w}^{Q\alpha}$ or $\mathcal{V}_{h,R}^{Q\alpha}$.

The associated graded with respect to $\mathfrak{W}$ is identified (as a commutative algebra) with the polynomial algebra generated by the (commutative version of the) affine currents and their derivatives, and the grading by $w$ simply counts the number of (arbitrarily differentiated) currents.

We will establish the key identity by making the following identifications:
\begin{eqnarray}
    m_0(L\delta+Q\alpha)+\sum_{\substack{p\geqslant2,\,q\geqslant1\\(p,q)=1}}m_{p/q}(L\delta+Q\alpha)&=& \sum_{w=0}^{L} w\,{\rm dim}V^{Q\alpha}_{L,w}~,\quad\text{\small(leading term identity)}\label{eq:leading_term_id_sl2}\\
    m_0(L\delta+Q\alpha) + \sum_{R=0}^{L}R\,\dim \mathcal{V}^{Q\alpha}_{L,R}&=&\sum_{w=0}^L w\,{\rm dim}V^{Q\alpha}_{L,w}~,\quad\text{\small(R-grading identity)}\label{eq:R_grading_id}
\end{eqnarray}
from which the identity \eqref{eq:sumR_vs_sumMpq} follows immediately.

To demonstrate \eqref{eq:leading_term_id_sl2}, observe that the highest power of $k$ in the two-point function of any operator with nonzero image in $V_{L,w}$ after passing to the associated graded is $k^w$. Thus, the determinant of the level-$L$ Gram matrix will produce a leading term of order $k^{\sum_{w}w\,\dim V_{L,w}}$. Compatibility with the Gorelik--Kac determinant formula \eqref{eq:KacDetAKM} then implies \eqref{eq:leading_term_id_sl2}.

To demonstrate \eqref{eq:R_grading_id}, we first recall the generating function of the critical-level exponents $m_0(L\delta)$ as given in \eqref{eq:M0answer}, which we rewrite here as a series in fugacities $q$ and $y$,
\begin{equation}
    F_0(q,y) \coloneqq \sum_{\nu\in\hat{\Delta}_+}m_0(\nu=L\delta+Q\alpha)q^Ly^Q = \frac{\sum_{i=2}^\infty \frac{q^i}{1-q^i}}{\prod_{j=1}^\infty (1-q^{j})(1-y q^{j})(1-y^{-1}q^j)}~.
\end{equation}
We now introduce additional generating functions,
\begin{eqnarray}
    F_\mathfrak{W}(q,y) &=& \sum_{L,Q}\left(\sum_{w=0}^L w\,{\rm dim}V^{Q\alpha}_{L,w}\right)q^Ly^Q~,\\
    F_\mathfrak{R}(q,y) &=& \sum_{L,Q}\left(\sum_{R=0}^L R\,{\rm dim}{\mathcal V}^{Q\alpha}_{L,R}\right)q^Ly^Q~.
\end{eqnarray}
The former of these is straightforward to express given the description of the associated graded as a polynomial algebra. We start with a relative of the vacuum Kostant partition function, with an extra fugacity measuring the weights of the generators with respect to the $w$ grading:
\begin{equation}\label{eq:fJ}
    f_\mathfrak{W}(q,y,w)\coloneqq \frac{1}{\prod_{j=1}^\infty (1-w\,q^{j})(1-w\,y\, q^{j})(1-w\,y^{-1}q^j)}~.
\end{equation}
Then for the generating function in question we differentiate by the parameter $w$ and evaluate at $w=1$,
\begin{equation}
    F_{\mathfrak{W}}(q,y)=\partial_w f_{\mathfrak W}(q,y,w)\big|_{w=1}=\frac{\left(\sum_{j=1}^\infty\frac{q^j}{1-q^j}+\sum_{j=1}^\infty\frac{yq^j}{1-yq^j}+\sum_{j=1}^\infty\frac{y^{-1}q^j}{1-y^{-1}q^j}\right)}{\prod_{j=1}^\infty (1-q^{j})(1-y q^{j})(1-y^{-1}q^j)}~.
\end{equation}
The generating function $F_\mathfrak{R}(q,y)$ can be found using a similar trick, but using the following analog of \eqref{eq:fJ},
\begin{equation}\label{eq:fR}
    f_{\mathfrak R}(q,y,r)=\frac{\prod_{n=2}^\infty (1-r^2\,q^{n})}{\prod_{n=2}^\infty (1-r\,q^{n})}\cdot\frac{1}{\prod_{j=1}^\infty (1-r\,q^{j})(1-r\,y\, q^{j})(1-r\,y^{-1}q^j)}~,
\end{equation}
where now $r$ is conjugate to the $R$-grading on the associated graded. The form of this expression can be understood by realizing $\rm{gr}^\mathfrak{R}V^k(\mathfrak{sl}_2)$ as a polynomial ring in the (commutative) currents and their derivatives subject to the constraint $J^aJ_a=0$ (giving the numerator), but with an additional generator associated with the stress tensor (which now has $R=1$, giving the first term in the denominator).

Computing the generating function proceeds analogously, and we have
\begin{equation}
    \begin{split}
    F_{\mathfrak R}(q,y)=\partial_r f_{\mathfrak R}(q,y,r)\big|_{r=1}&=\partial_t f_{\mathfrak W}(q,y,t)\big|_{t=1}+\partial_r \left(\frac{\prod_{n=2}^\infty (1-r^2\,q^{n})}{\prod_{n=2}^\infty (1-r\,q^{n})}\right)\bigg|_{r=1} f_{\mathfrak{W}}(q,y,1)\\
    &=F_{\mathfrak W}(q,y)-F_0(q,y)~.
    \end{split}
\end{equation}
We conclude that $F_{\mathfrak R}(q,y)+F_0(q,y)=F_{\mathfrak W}(q,y)$, which establishes \eqref{eq:R_grading_id}.

\subsection{Exponent identity for \texorpdfstring{$\mathfrak{sl}_N$}{sl(N)}}

Here we derive the generalization of the previous identities to the case of $\mathfrak{g}=\mathfrak{sl}_N$. We again define generating functions
\begin{eqnarray}
    F_\mathfrak{W}(q,\vec{y}) &=& \sum_{L,\nu}\left(\sum_{w=0}^L w\,{\rm dim}V^{(\nu)}_{L,w}\right)q^Ly^\nu~,\\
    F_\mathfrak{R}(q,\vec{y}) &=& \sum_{L,\nu}\left(\sum_{R=0}^L R\,{\rm dim}{\mathcal V}^{(\nu)}_{L,R}\right)q^Ly^\nu~.
\end{eqnarray}
The analogue of the leading term identity \eqref{eq:leading_term_id_sl2} is as follows, and is proven precisely as before by comparing overall powers of $k$ in the determinant,
\begin{equation}\label{eq:leading_term_id_sln}
    \sum_{\substack{p\geqslant2,\,q\geqslant1\\(p,q)=1}}m_{p/q}(L\delta+\nu)+m_0(L\delta+\nu) = \sum_{w=0}^{L} w\,{\rm dim}V^{(\nu)}_{L,w}~,
\end{equation}
where now $\nu$ denotes a general element of the root lattice of $\mathfrak{g}$ and again $V_{L,w}^{(\nu)}$ is the charge-$\nu$ sector of the associated graded of $V^{k}(\mathfrak{g})$ of conformal level $L$ and weight $w$.

It remains to generalize \eqref{eq:R_grading_id}. We begin with the generating function for the multiplicities of critical zeros. For $\mathfrak{sl}_N$ we have $s_{m-1}-s_{m}=1$ for $m=2,\dots,N$, so \eqref{eq:M0answerGeneral} (now written with $q$ as a fugacity for the imaginary root) specializes as
\begin{equation}\label{eq:F0}
    F_0(q)=\frac{\sum_{m=2}^{N}\sum_{i=m}^\infty \frac{q^i}{1-q^i}}{\tilde{R}(q)}~,\qquad \tilde{R}(q)\coloneqq \prod_{i=1}^\infty\prod_{\alpha\in\Delta}(1-q^ie^{-\alpha})^{\dim\mathfrak{g}_\alpha}~.
\end{equation}
The analog of \eqref{eq:fJ}, defined with respect to the conformal weight-based filtration, is now 
\begin{equation}
    f_\mathfrak{W}(q,w)=\frac{1}{\prod_{i=1}^\infty\prod_{\alpha\in\Delta}(1-wq^i\,e^{-\alpha})^{\dim\mathfrak{g}_{\alpha}}}~,
\end{equation}
while the analog of \eqref{eq:fR}, now defined with respect to the Casimir weight-based filtration (the postulated $\mathfrak{R}$-filtration) is given by
\begin{equation}\label{eq:fR_sl3}
    f_{\mathfrak{R}}(q,w)=\frac{\prod_{i=2}^N\prod_{n=i}^\infty(1-w^i q^{n})}{\prod_{i=2}^N\prod_{n=i}^\infty(1-w^{i-1}q^n)}\frac{1}{\prod_{i=1}^\infty\prod_{\alpha\in\Delta}(1-wq^ie^{-\alpha})^{\dim\mathfrak{g}_\alpha}}~.
\end{equation}
We can then verify the identity
\begin{equation}\label{eq:F0=FJ-FR}
\begin{split}
    F_\mathfrak{R}(q)\coloneqq\partial_w f_\mathfrak{R}(q,w)\big|_{w=1}&=\partial_w f_\mathfrak{W}(q,w)\big|_{w=1}+\partial_w \left(\frac{\prod_{i=2}^N\prod_{n=i}^\infty(1-w^i q^{n})}{\prod_{i=2}^N\prod_{n=i}^\infty(1-w^{i-1}q^n)}\right)\bigg|_{t=1}\,\frac{1}{\tilde{R}(q)}\\
    &=F_\mathfrak{R}(q)-F_0(q)~.
\end{split}
\end{equation}
We conclude that $F_\mathfrak{R}(q)+F_0(q)=F_\mathfrak{W}(q)$, establishing \eqref{eq:R_grading_id} for $\mathfrak{sl}_N$.

\subsection{Exponent identity for all simply-laced \texorpdfstring{$\mathfrak{g}$}{g}}

Next consider $\mathfrak g$ of type $D$, namely $\mathfrak g=\mathfrak{so}_{2N}.$ There are Casimirs of order $m=2,4,\dots,2N-2$, which together with a Pfaffian of order $m=N$ form an algebraically independent set. (See Corollary~5.4.6 in \cite{molev2018sugawara}, or Section~11 in \cite{bincer2013lie}.) For notational convenience, let us define the set of degrees of independent Casimirs $\mathcal{M}_N\coloneqq\{2,4,\dots,2N-2,N\}$, where the multiplicity of $N$ is $2$ if $N$ is even and $1$ if $N$ is odd. The $\mathfrak{so}_{2N}$ analog of \eqref{eq:F0} can then be written as
\begin{equation}
    F_0(q)=\frac{\sum_{m\in\mathcal{M}_N}\sum_{i=m}^\infty \frac{q^i}{1-q^i}}{\tilde R(q)}~.
\end{equation}
It can be checked that $\mathfrak{so}_{2N}$ one has $s_{m-1}-s_{m}=1$ for $m\in\mathcal{M}_N$ and zero otherwise for $N$ odd, and if $N$ is even then $s_{m-1}-s_{m}=2$ for $m=N$ and the rest the same. Thus, the above generating function matches the one in \eqref{eq:M0answerGeneral}. A calculation similar to \eqref{eq:F0=FJ-FR} then establishes the analogous identity \eqref{eq:R_grading_id} for $\mathfrak{g}$ of type $D$.

To further generalize the identity to $\mathfrak{g}$ of type $E$, one needs only that $s_{m-1}-s_{m}$ coincides with the number of order-$m$ Casimirs in $E_{6,7,8}$ as well, which can be confirmed by direct examination.

Finally, the generalization of \eqref{eq:M0answerGeneral} to nonsimply-laced $\mathfrak{g}$ holds where $s_j$ is defined as the number of positive roots of height $j$, namely $s_j\coloneqq\# \{\alpha\in \Delta^+: \frac{2(\rho|\alpha)}{(\alpha|\alpha)}=j\}$. The identity \eqref{eq:R_grading_id} then follows from a calculation of type \eqref{eq:F0=FJ-FR} upon identifying $s_{m-1}-s_{m}$ with the number of order-$m$ Casimirs in this general case. This indeed holds (see, \emph{e.g.}, Sections~1 and 9 in~\cite{Kostant}).


\bibliographystyle{aux/ytphys}
\bibliography{aux/refs}

@article{Liendo:2015ofa,
    author = "Liendo, Pedro and Ramirez, Israel and Seo, Jihye",
    title = "{Stress-tensor OPE in $ \mathcal{N}=2 $ superconformal theories}",
    eprint = "1509.00033",
    archivePrefix = "arXiv",
    primaryClass = "hep-th",
    reportNumber = "HU-EP-15-39, DESY-15-164",
    doi = "10.1007/JHEP02(2016)019",
    journal = "JHEP",
    volume = "02",
    pages = "019",
    year = "2016"
}

@article{arakawa2021arc,
  title={Arc spaces and vertex algebras},
  author={Arakawa, Tomoyuki and Moreau, Anne},
  journal={Springer Monograph, April},
  volume={17},
  pages={170},
  year={2021}
}

@article{Foda:2019guo,
    author = "Foda, Omar and Zhu, Rui-Dong",
    title = "{Closed form fermionic expressions for the Macdonald index}",
    eprint = "1912.01896",
    archivePrefix = "arXiv",
    primaryClass = "hep-th",
    reportNumber = "DIAS-STP-19-08",
    doi = "10.1007/JHEP06(2020)157",
    journal = "JHEP",
    volume = "06",
    pages = "157",
    year = "2020"
}

@article{Xie:2016evu,
    author = "Xie, Dan and Yan, Wenbin and Yau, Shing-Tung",
    title = "{Chiral algebra of the Argyres-Douglas theory from M5 branes}",
    eprint = "1604.02155",
    archivePrefix = "arXiv",
    primaryClass = "hep-th",
    doi = "10.1103/PhysRevD.103.065003",
    journal = "Phys. Rev. D",
    volume = "103",
    number = "6",
    pages = "065003",
    year = "2021"
}

@article{Gadde:2011uv,
    author = "Gadde, Abhijit and Rastelli, Leonardo and Razamat, Shlomo S. and Yan, Wenbin",
    title = "{Gauge Theories and Macdonald Polynomials}",
    eprint = "1110.3740",
    archivePrefix = "arXiv",
    primaryClass = "hep-th",
    reportNumber = "YITP-SB-11-30",
    doi = "10.1007/s00220-012-1607-8",
    journal = "Commun. Math. Phys.",
    volume = "319",
    pages = "147--193",
    year = "2013"
}

@article{Cordova:2015nma,
    author = "Cordova, Clay and Shao, Shu-Heng",
    title = "{Schur Indices, BPS Particles, and Argyres-Douglas Theories}",
    eprint = "1506.00265",
    archivePrefix = "arXiv",
    primaryClass = "hep-th",
    doi = "10.1007/JHEP01(2016)040",
    journal = "JHEP",
    volume = "01",
    pages = "040",
    year = "2016"
}

@article{Kapustin:2006hi,
    author = "Kapustin, Anton",
    title = "{Holomorphic reduction of N=2 gauge theories, Wilson-'t Hooft operators, and S-duality}",
    eprint = "hep-th/0612119",
    archivePrefix = "arXiv",
    reportNumber = "CALT-68-2623",
    month = "12",
    year = "2006"
}

@article{Oh:2019mcg,
    author = "Oh, Jihwan and Yagi, Junya",
    title = "{Poisson vertex algebras in supersymmetric field theories}",
    eprint = "1908.05791",
    archivePrefix = "arXiv",
    primaryClass = "hep-th",
    doi = "10.1007/s11005-020-01290-0",
    journal = "Lett. Math. Phys.",
    volume = "110",
    number = "8",
    pages = "2245--2275",
    year = "2020"
}

@article{Kostant,
 ISSN = {00029327, 10806377},
 URL = {http://www.jstor.org/stable/2372999},
 author = {Bertram Kostant},
 journal = {American Journal of Mathematics},
 number = {4},
 pages = {973--1032},
 publisher = {Johns Hopkins University Press},
 title = {The Principal Three-Dimensional Subgroup and the Betti Numbers of a Complex Simple Lie Group},
 urldate = {2024-04-15},
 volume = {81},
 year = {1959}
}

@book{bincer2013lie,
  title={Lie Groups and Lie Algebras-A Physicist's Perspective},
  author={Bincer, Adam M},
  year={2013},
  publisher={Oxford University Press, USA}
}

@book{molev2018sugawara,
  title={Sugawara operators for classical Lie algebras},
  author={Molev, Alexander},
  volume={229},
  year={2018},
  publisher={American Mathematical Soc.}
}

@article{Bais:1987dc,
    author = "Bais, F. A. and Bouwknegt, P. and Surridge, M. and Schoutens, K.",
    title = "{Extensions of the Virasoro Algebra Constructed from Kac-Moody Algebras Using Higher Order Casimir Invariants}",
    reportNumber = "ITFA-87-12, THU-87-18",
    doi = "10.1016/0550-3213(88)90631-1",
    journal = "Nucl. Phys. B",
    volume = "304",
    pages = "348--370",
    year = "1988"
}

@article{Bouwknegt:1992wg,
    author = "Bouwknegt, Peter and Schoutens, Kareljan",
    title = "{W symmetry in conformal field theory}",
    eprint = "hep-th/9210010",
    archivePrefix = "arXiv",
    reportNumber = "CERN-TH-6583-92, ITPO-SB-92-23",
    doi = "10.1016/0370-1573(93)90111-P",
    journal = "Phys. Rept.",
    volume = "223",
    pages = "183--276",
    year = "1993"
}

@article{Dolan:2002zh,
    author = "Dolan, F. A. and Osborn, H.",
    title = "{On short and semi-short representations for four-dimensional superconformal symmetry}",
    eprint = "hep-th/0209056",
    archivePrefix = "arXiv",
    reportNumber = "DAMTP-02-114",
    doi = "10.1016/S0003-4916(03)00074-5",
    journal = "Annals Phys.",
    volume = "307",
    pages = "41--89",
    year = "2003"
}

@article{Bianchi:2019sxz,
    author = "Bianchi, Lorenzo and Lemos, Madalena",
    title = "{Superconformal surfaces in four dimensions}",
    eprint = "1911.05082",
    archivePrefix = "arXiv",
    primaryClass = "hep-th",
    reportNumber = "CERN-TH-2019-190",
    doi = "10.1007/JHEP06(2020)056",
    journal = "JHEP",
    volume = "06",
    pages = "056",
    year = "2020"
}

@article{Lemos:2015orc,
    author = "Lemos, Madalena and Liendo, Pedro",
    title = "{$\mathcal{N}=2$ central charge bounds from $2d$ chiral algebras}",
    eprint = "1511.07449",
    archivePrefix = "arXiv",
    primaryClass = "hep-th",
    reportNumber = "DESY-15-230, HU-EP-15-56, HU-EP-15/56",
    doi = "10.1007/JHEP04(2016)004",
    journal = "JHEP",
    volume = "04",
    pages = "004",
    year = "2016"
}

@article{Nirschl:2004pa,
    author = "Nirschl, M. and Osborn, H.",
    title = "{Superconformal Ward identities and their solution}",
    eprint = "hep-th/0407060",
    archivePrefix = "arXiv",
    reportNumber = "DAMTP-04-51",
    doi = "10.1016/j.nuclphysb.2005.01.013",
    journal = "Nucl. Phys. B",
    volume = "711",
    pages = "409--479",
    year = "2005"
}

@article{Maldacena:2011jn,
    author = "Maldacena, Juan and Zhiboedov, Alexander",
    title = "{Constraining Conformal Field Theories with A Higher Spin Symmetry}",
    eprint = "1112.1016",
    archivePrefix = "arXiv",
    primaryClass = "hep-th",
    doi = "10.1088/1751-8113/46/21/214011",
    journal = "J. Phys. A",
    volume = "46",
    pages = "214011",
    year = "2013"
}

@article{Alba:2013yda,
    author = "Alba, Vasyl and Diab, Kenan",
    title = "{Constraining conformal field theories with a higher spin symmetry in d=4}",
    eprint = "1307.8092",
    archivePrefix = "arXiv",
    primaryClass = "hep-th",
    month = "7",
    year = "2013"
}

@inproceedings{Simmons-Duffin:2016gjk,
    author = "Simmons-Duffin, David",
    title = "{The Conformal Bootstrap}",
    booktitle = "{Theoretical Advanced Study Institute in Elementary Particle Physics}: {New Frontiers in Fields and Strings}",
    eprint = "1602.07982",
    archivePrefix = "arXiv",
    primaryClass = "hep-th",
    doi = "10.1142/9789813149441_0001",
    pages = "1--74",
    year = "2017"
}

@article{dong2014unitary,
  title={Unitary vertex operator algebras},
  author={Dong, Chongying and Lin, Xingjun},
  journal={Journal of algebra},
  volume={397},
  pages={252--277},
  year={2014},
  publisher={Elsevier}
}

@article{Beem:2016cbd,
    author = "Beem, Christopher and Peelaers, Wolfger and Rastelli, Leonardo",
    title = "{Deformation quantization and superconformal symmetry in three dimensions}",
    eprint = "1601.05378",
    archivePrefix = "arXiv",
    primaryClass = "hep-th",
    doi = "10.1007/s00220-017-2845-6",
    journal = "Commun. Math. Phys.",
    volume = "354",
    number = "1",
    pages = "345--392",
    year = "2017"
}

@article{Etingof:2019guc,
    author = "Etingof, Pavel and Stryker, Douglas",
    title = "{Short Star-Products for Filtered Quantizations, I}",
    eprint = "1909.13588",
    archivePrefix = "arXiv",
    primaryClass = "math.RT",
    doi = "10.3842/SIGMA.2020.014",
    journal = "SIGMA",
    volume = "16",
    pages = "014",
    year = "2020"
}

@article{Etingof:2020fls,
    author = "Etingof, Pavel and Klyuev, Daniil and Rains, Eric and Stryker, Douglas",
    title = "{Twisted Traces and Positive Forms on Quantized Kleinian Singularities of Type A}",
    eprint = "2009.09437",
    archivePrefix = "arXiv",
    primaryClass = "math.QA",
    doi = "10.3842/SIGMA.2021.029",
    journal = "SIGMA",
    volume = "17",
    pages = "029",
    year = "2021"
}

@article{Feigin:1982tg,
    author = "Feigin, B. L. and Fuks, D. B.",
    title = "{Verma modules over the Virasoro algebra}",
    doi = "10.1007/BF01083148",
    journal = "Funct. Anal. Appl.",
    volume = "17",
    pages = "241--241",
    year = "1983"
}

@article{gorelik2007simplicity,
  title={On simplicity of vacuum modules},
  author={Gorelik, Maria and Kac, Victor},
  journal={Advances in Mathematics},
  volume={211},
  number={2},
  pages={621--677},
  year={2007},
  publisher={Elsevier}
}

@article{Beem:2019tfp,
    author          = "Beem, Christopher and Meneghelli, Carlo and Rastelli, Leonardo",
    title           = "{Free Field Realizations from the Higgs Branch}",
    eprint          = "1903.07624",
    archivePrefix   = "arXiv",
    primaryClass    = "hep-th",
    doi             = "10.1007/JHEP09(2019)058",
    journal         = "JHEP",
    volume          = "09",
    pages           = "058",
    year            = "2019"
}

@article{Beem:2021jnm,
    author          = "Beem, Christopher and Meneghelli, Carlo",
    title           = "{Geometric free field realization for the genus-two class S theory of type a1}",
    eprint          = "2104.11668",
    archivePrefix   = "arXiv",
    primaryClass    = "hep-th",
    doi             = "10.1103/PhysRevD.104.065015",
    journal         = "Phys. Rev. D",
    volume          = "104",
    number          = "6",
    pages           = "065015",
    year            = "2021"
}

@article{Beem:2019snk,
    author          = "Beem, Christopher and Meneghelli, Carlo and Peelaers, Wolfger and Rastelli, Leonardo",
    title           = "{VOAs and rank-two instanton SCFTs}",
    eprint          = "1907.08629",
    archivePrefix   = "arXiv",
    primaryClass    = "hep-th",
    reportNumber    = "YITP-SB-19-20",
    doi             = "10.1007/s00220-020-03746-9",
    journal         = "Commun. Math. Phys.",
    volume          = "377",
    number          = "3",
    pages           = "2553--2578",
    year            = "2020"
}

@article{Dedushenko:2019mzv,
    author = "Dedushenko, Mykola",
    title = "{From VOAs to short star products in SCFT}",
    eprint = "1911.05741",
    archivePrefix = "arXiv",
    primaryClass = "hep-th",
    reportNumber = "CALT-TH 2019-040, CALT-TH-2019-040",
    doi = "10.1007/s00220-021-04066-2",
    journal = "Commun. Math. Phys.",
    volume = "384",
    number = "1",
    pages = "245--277",
    year = "2021"
}

@article{Song:2016yfd,
    author = "Song, Jaewon",
    title = "{Macdonald Index and Chiral Algebra}",
    eprint = "1612.08956",
    archivePrefix = "arXiv",
    primaryClass = "hep-th",
    doi = "10.1007/JHEP08(2017)044",
    journal = "JHEP",
    volume = "08",
    pages = "044",
    year = "2017"
}

@article{moriwaki2020classification,
  title={On classification of conformal vectors in vertex operator algebra and the vertex algebra automorphism group},
  author={Moriwaki, Yuto},
  journal={Journal of Algebra},
  volume={546},
  pages={689--702},
  year={2020},
  publisher={Elsevier}
}

@article{arakawa2024generalized,
  title         = {Generalized Grothendieck's simultaneous resolution and associated varieties of simple affine vertex algebras},
  author        = {Arakawa, Tomoyuki and Futorny, Vyacheslav and Krizka, Libor},
  journal       = {arXiv e-prints},
  eprint        = {2404.02365},
  eid           = {arXiv:2404.02365},
  pages         = {arXiv:2404.02365},
  archivePrefix = {arXiv},
  primaryClass  = {math.RT},
  year          = {2024}
}

@article{Arakawa:2010ni,
	author         = {{Arakawa}, Tomoyuki},
	title          = "{Associated varieties of modules over Kac-Moody algebras and $C_2$-cofiniteness of W-algebras}",
	journal        = {arXiv e-prints},
	year           = {2010},
	eid            = {arXiv:1004.1554},
	pages          = {arXiv:1004.1554},
	archivePrefix  = {arXiv},
	eprint         = {1004.1554},
	primaryClass   = {math.QA},
}

@article{Arakawa_Moreau:2017157,
title = {Sheets and associated varieties of affine vertex algebras},
journal = {Advances in Mathematics},
volume = {320},
pages = {157-209},
year = {2017},
issn = {0001-8708},
doi = {https://doi.org/10.1016/j.aim.2017.08.039},
url = {https://www.sciencedirect.com/science/article/pii/S0001870817302311},
author = {Tomoyuki Arakawa and Anne Moreau}
}

@article{li2004vertex,
  title={Vertex algebras and vertex Poisson algebras},
  author={Li, Haisheng},
  journal={Communications in Contemporary Mathematics},
  volume={6},
  number={01},
  pages={61--110},
  year={2004},
  publisher={World Scientific}
}

@article{Argyres:2022npi,
    author = "Argyres, Philip C. and Lotito, Matteo and Weaver, Mitch",
    title = "{Vertex algebra of extended operators in 4d N=2 superconformal field theories. Part I}",
    eprint = "2211.04410",
    archivePrefix = "arXiv",
    primaryClass = "hep-th",
    doi = "10.1007/JHEP10(2023)175",
    journal = "JHEP",
    volume = "10",
    pages = "175",
    year = "2023"
}

@article{Pan:2020cgc,
    author = "Pan, Yiwen and Peelaers, Wolfger",
    title = "{Deformation quantizations from vertex operator algebras}",
    eprint = "1911.09631",
    archivePrefix = "arXiv",
    primaryClass = "hep-th",
    doi = "10.1007/JHEP06(2020)127",
    journal = "JHEP",
    volume = "06",
    pages = "127",
    year = "2020"
}

@article{Beem:2013sza,
    author         = "Beem, Christopher and Lemos, Madalena and Liendo, Pedro and Peelaers, Wolfger and Rastelli, Leonardo and van Rees, Balt C.",
    title          = "{Infinite Chiral Symmetry in Four Dimensions}",
    journal        = "Commun. Math. Phys.",
    volume         = "336",
    year           = "2015",
    number         = "3",
    pages          = "1359-1433",
    doi            = "10.1007/s00220-014-2272-x",
    eprint         = "1312.5344",
    archivePrefix  = "arXiv",
    primaryClass   = "hep-th",
    reportNumber   = "YITP-SB-13-45, CERN-PH-TH-2013-311, HU-EP-13-78",
    SLACcitation   = "%%CITATION = ARXIV:1312.5344;%%"
}

@article{Beem:2018duj,
    author = "Beem, Christopher",
    title = "{Flavor Symmetries and Unitarity Bounds in ${\mathcal N}=2$ Superconformal Field Theories}",
    eprint = "1812.06099",
    archivePrefix = "arXiv",
    primaryClass = "hep-th",
    doi = "10.1103/PhysRevLett.122.241603",
    journal = "Phys. Rev. Lett.",
    volume = "122",
    number = "24",
    pages = "241603",
    year = "2019"
}

@article{Beem:2017ooy,
	author         = "Beem, Christopher and Rastelli, Leonardo",
	title          = "{Vertex operator algebras, Higgs branches, and modular differential equations}",
	eprint         = "1707.07679",
	archivePrefix  = "arXiv",
	primaryClass   = "hep-th",
	reportNumber   = "YITP-SB-17-27",
	doi            = "10.1007/JHEP08(2018)114",
	journal        = "JHEP",
	volume         = "08",
	pages          = "114",
	year           = "2018"
}

@article{Beem:2024fom,
    author = "Beem, Christopher and Deb, Anirudh and Martone, Mario and Meneghelli, Carlo and Rastelli, Leonardo",
    title = "{Free field realizations for rank-one SCFTs}",
    eprint = "2407.01674",
    archivePrefix = "arXiv",
    primaryClass = "hep-th",
    doi = "10.1007/JHEP12(2024)004",
    journal = "JHEP",
    volume = "12",
    pages = "004",
    year = "2024"
}

@article{Jeong:2019pzg,
    author = "Jeong, Saebyeok",
    title = "{SCFT/VOA correspondence via $\Omega$-deformation}",
    eprint = "1904.00927",
    archivePrefix = "arXiv",
    primaryClass = "hep-th",
    reportNumber = "YITP-SB-19-7",
    doi = "10.1007/JHEP10(2019)171",
    journal = "JHEP",
    volume = "10",
    pages = "171",
    year = "2019"
}

@article{Oh:2019bgz,
    author = "Oh, Jihwan and Yagi, Junya",
    title = "{Chiral algebras from $\Omega$-deformation}",
    eprint = "1903.11123",
    archivePrefix = "arXiv",
    primaryClass = "hep-th",
    doi = "10.1007/JHEP08(2019)143",
    journal = "JHEP",
    volume = "08",
    pages = "143",
    year = "2019"
}

@article{BeemGarner:Hodge,
    author = "Beem, Christopher and Garner, Niklas",
    title = "{On the semi-infinite cohomology of graded-unitary vertex algebras}",
    eprint = "2509.10364",
    archivePrefix = "arXiv",
    primaryClass = "math.QA",
    month = "9",
    year = "2025"
}

@unpublished{Beem:String-Math-2017,
title= "Comments on vertex algebras for $\mathcal{N}=2$ {SCFTs}",
author = {Beem, Christopher},
year = {2017},
month = {July},
note= {String Math 2017},
}

@unpublished{Arakawa_talks,
title={R-filtration and Vogan filtration},
author = {Arakawa, Tomoyuki},
year    =   {2021},
month   =   {October},
note    =   {Pure Spinors, Superalgebras, and Holomorphic Twists - Heidelberg University},
}

@unpublished{Beem:Pollica,
title= "4d $\mathcal{N}=2$ {SCFTs} and {VOAs}",
author = {Beem, Christopher},
year = {2019},
month = {June},
note= {Pollica Summer
  Workshop "Mathematical and Geometric Tools for Conformal Field Theories,"},
}

@unpublished{Beem:String-Math-2019,
title= "Building VOAs out of Higgs branches",
author = {Beem, Christopher},
year = {2019},
month = {July},
note= {String Math 2019},
}

@incollection{Arakawa2018,
author="Arakawa, Tomoyuki
and Kawasetsu, Kazuya",
editor="Kac, Victor G.
and Popov, Vladimir L.",
title="Quasi-lisse Vertex Algebras and Modular Linear Differential Equations",
bookTitle="Lie Groups, Geometry, and Representation Theory: A Tribute to the Life and Work of Bertram Kostant",
year="2018",
publisher="Springer International Publishing",
address="Cham",
pages="41--57",
isbn="978-3-030-02191-7",
doi="10.1007/978-3-030-02191-7_2",
url="https://doi.org/10.1007/978-3-030-02191-7_2"
}

\end{document}